\documentclass[
  journal=pasa,
  manuscript=research-paper,
  year=2024,
  volume=37,
]{cup-journal}

\usepackage{amsmath}
\usepackage[nopatch]{microtype}
\usepackage{booktabs}

\usepackage{lscape}

\usepackage{float,xcolor}
\usepackage[T1]{fontenc}

\usepackage{graphicx}	
\usepackage{amsmath}	
\usepackage{float}
 \usepackage{amssymb}	
\usepackage{geometry}
\usepackage{hyphenat}
\usepackage{mwe}
\usepackage{blindtext}
\usepackage{wrapfig}
\usepackage[figuresright]{rotating}
\usepackage{gensymb}
\usepackage{natbib}
\usepackage{tablefootnote}

\usepackage{caption}
\DeclareCaptionFormat{bold}{\textbf{#1 #2 }#3}
\DeclareCaptionLabelSeparator{dot}{.}
\usepackage{svg}
\usepackage{array}
\usepackage{booktabs}

\newcommand{\Msun}{\mbox{\,$\rm M_{\odot}$}}
\newcommand{\Lsun}{\mbox{\,$\rm L_{\odot}$}}
\newcommand{\Lsed}{\mbox{\,$\rm L_{SED}$}}
\newcommand{\Lplc}{\mbox{\,$\rm L_{PLC}$}}

\newcommand{\Porb}{\mbox{\,$\rm P_{Orb}$}}
 
\newcommand{\Teff}{$T_{\rm eff}$}

\newcommand{\logg}{$\log g$}
\newcommand{\mv}{$\xi_{\rm t}$}
\newcommand{\co}{$\textrm{C/O}$}
\newcommand{\cfe}{$\textrm{[C/Fe]}$}
\newcommand{\ofe}{$\textrm{[O/Fe]}$}

\newcommand{\sfe}{$\textrm{[s/Fe]}$}

\newcommand{\feh}{$\textrm{[Fe/H]}$}
\newcommand{\xfe}{$\textrm{[X/Fe]}$}
\newcommand{\xh}{$\textrm{[X/H]}$}

\newcommand{\kms}{km\,s$^{-1}$}

\newcommand{\sprocess}{\textit{s}-process }
\newcommand{\logepsilon}{$\log\varepsilon$}
\newcommand{\logepsilonsun}{$\log\varepsilon_{\odot}$}
\newcommand{\sigmatotal}{$\sigma_{\rm tot}$}
\newcommand{\sigmalinetoline}{$\sigma_{\rm |121|}$}

\newcommand{\Tcondensation}{$T_{\rm condensation}$}
\newcommand{\znti}{$\textrm{[Zn/Ti]}$}

\usepackage{subcaption}
\usepackage{setspace,hyperref}
\hypersetup{colorlinks=true,
            linkcolor=blue,
            urlcolor=blue,
            linktoc=all,
            citecolor=blue}

\title{\sprocess Enriched Evolved Binaries in the Galaxy and the Magellanic Clouds}

\author{Meghna Menon}
\affiliation{School of Mathematical and Physical Sciences, Macquarie University, Balaclava Road, Sydney, NSW 2109, Australia}
\alsoaffiliation{Astrophysics and Space Technologies Research Centre, Macquarie University, Balaclava Road, Sydney, NSW 2109, Australia}
\email[Meghna Menon]{meghnamukesh.menon1@students.mq.edu.au}

\author{Devika Kamath}
\affiliation{School of Mathematical and Physical Sciences, Macquarie University, Balaclava Road, Sydney, NSW 2109, Australia}
\alsoaffiliation{Astrophysics and Space Technologies Research Centre, Macquarie University, Balaclava Road, Sydney, NSW 2109, Australia}
\alsoaffiliation{INAF, Observatory of Rome, Via Frascati 33, I-00077 Monte Porzio Catone (RM), Italy}

\author{Maksym Mohorian}
\affiliation{School of Mathematical and Physical Sciences, Macquarie University, Balaclava Road, Sydney, NSW 2109, Australia}
\alsoaffiliation{Astrophysics and Space Technologies Research Centre, Macquarie University, Balaclava Road, Sydney, NSW 2109, Australia}

\author{Hans Van Winckel}
\affiliation{Instituut voor Sterrenkunde, K.U.Leuven, Celestijnenlaan 200D bus 2401, B-3001, Leuven, Belgium}

\author{Paolo Ventura}
\affiliation{INAF, Observatory of Rome, Via Frascati 33, I-00077 Monte Porzio Catone (RM), Italy}

\keywords{stars: evolution - binaries - AGB and post-AGB - chemically peculiar - abundances, techniques: spectroscopic, galaxies: Magellanic Clouds} 

\begin{document}

\begin{abstract}
Post-asymptotic giant branch stars (post-AGB) in binary systems, with typical orbital periods between $\sim\!100$ to $\sim\!1000$ days, result from a poorly understood interaction that terminates their precursory AGB phase. The majority of these binaries display a photospheric anomaly called ‘chemical depletion’, thought to arise from an interaction between the circumbinary disc and the post-AGB star, leading to the reaccretion of pure gas onto the star, devoid of refractory elements due to dust formation. In this paper, we focus on a subset of chemically peculiar binary post-AGBs in the Galaxy and the Magellanic Clouds (MCs). Our detailed stellar parameter and chemical abundance analysis utilising high-resolution optical spectra from VLT+UVES revealed that our targets span a \Teff\ of 4900\,-\,7250\,K and \feh\ of -0.5\,-\, -1.57\,dex. Interestingly, these targets exhibit a carbon (\cfe\ ranging from 0.5\,-\,1.0 dex, dependant on metallicity) and \sprocess enrichment ($\sfe\,\geq\!1$dex) contrary to the commonly observed chemical depletion pattern. Using spectral energy distribution (SED) fitting and period-luminosity-colour (PLC) relation methods, we determine the luminosity of the targets (2700\,$-$\,8300\Lsun), which enables confirmation of their evolutionary phase and estimation of initial masses (as a function of metallicity) (1\,-\,2.5\Msun). In conjunction with predictions from dedicated ATON stellar evolutionary models, our results indicate a predominant intrinsic enrichment of carbon and \sprocess elements in our binary post-AGB targets. We qualitatively rule out extrinsic enrichment and inherited \sprocess enrichment from the host galaxy as plausible explanations for the observed overabundances. Our chemically peculiar subset of intrinsic carbon and \sprocess enriched binary post-AGBs also hints at potential variation in the efficiency of chemical depletion between stars with C-rich and O-rich circumbinary disc chemistries. However, critical observational studies of circumbinary disc chemistry, along with specific condensation temperature estimates in C-rich environments, are necessary to address gaps in our current understanding of disc-binary interactions inducing chemical depletion in binary post-AGB systems.
\end{abstract}


\section{Introduction}
\label{sec:intro}
Standard stellar evolutionary models, describing the evolution of low- and intermediate-mass (LIM, 1-8\Msun) single stars, predict the occurrence of the slow neutron capture process (\sprocess) during their Asymptotic Giant Branch (AGB) phase \citep[see][]{Maurizio2001,Herwig2005,karakas14a}. The \sprocess results in the production of approximately half of the elements heavier than iron \citep[see][]{Gallino1998,Kappeler2011,kobayashi20}. The other half is thought to be due to the rapid neutron capture process (\textit{r}-process) happening in massive stars ($>\!8\Msun$) \citep[see][]{Wanajo2011,Kajino2019}. Recent investigations suggest the existence of additional nucleosynthetic processes such as the intermediate neutron capture process \citep[\textit{i}-process, see][]{Choplin2021,Choplin2022,Choplin2023}, contributing to the production of heavy elements in low-mass (1-3\Msun) AGB stars. The validity of these predictions has been observationally verified through studies of AGB stars \citep[e.g.,][]{Abia2002,Jonsell2006,Straniero2014,Cseh2019,Shetye2020} and post-AGB stars \citep[e.g.,][]{Vanwinckel2000,Reyniers2003,Kamath2014,kamath15}. 

Post-AGB stars serve as superior tracers of AGB nucleosynthesis and dredge-up compared to AGB stars. \citep[see][and references therein]{vanwinckel03,kamath20,Kamath2022Universe}. Notably, their higher photospheric temperatures ($\sim\!3000$ K to $\sim\!30000$ K) compared to their preceding AGB phase ($\sim\!2000$ K to $\sim\!4000$ K) ensure that the optical spectra of post-AGB stars allow for the detection of atomic transitions of a wide range of elements from CNO up to the heaviest \sprocess elements, well beyond the Ba peak \citep[e.g.,][and references therein]{Vanwinckel2000,desmedt12,desmedt2016}.

Post-AGB stars exhibit a distinctive Infrared (IR) excess, indicative of their dusty circumstellar environment \citep{vanwinckel03}. Detailed investigations of the spectral energy distribution (SED) in optically bright post-AGB stars have revealed two main categories based on the nature of their IR excess: ‘shell-source' and ‘disc-source' \citep[][and references therein]{vanwinckel03,Gezer2015,Manick2018}. The shell-sources represent single post-AGB stars with a distinctive double-peaked SED. The first peak (near-IR) corresponds to the photospheric component, while the second peak (mid-IR) characterises a detached and expanding shell of cold dust—an AGB dust shell remnant. However, the disc-sources display a broad IR excess, commencing in the near-IR region, indicating the existence of a stable compact dusty disc \citep[see][for more details on SED classifications]{Gezer2015}. Observational studies have now firmly established that the presence of the stable compact disc (circumbinary disc) around the disc-sources is attributed to their binary nature \citep[e.g.,][]{vanwinckel09,oomen18,Kluska2022}.

Observational studies of post-AGB stars in the Galaxy \citep[e.g.,][]{Vanwinckel2000,rao12,deSmedt2015,Kamath2022} and the Magellanic Clouds (MCs) \citep[e.g.,][]{Kamath2014,kamath15} have shown that they are far more chemically diverse than anticipated. Typically, low-mass single post-AGB stars are carbon and \sprocess enriched, a reflection of the nucleosynthetic processes that occur during and prior to their AGB phase \citep[eg][]{Reyniers2003}. Notably, some single post-AGB stars (shell-sources) stand out as the most \sprocess enriched objects known to date \citep[e.g.,][in the Galaxy and the MCs respectively]{Reyniers2004,desmedt12}. By and large, theoretical single-star low-mass AGB models are in agreement with observations \citep[e.g.,][]{vanaarle11,Desmedth2014,deSmedt2015,desmedt2016}. However, a study by \citet{kamath17} reported a subset of luminous single post-AGB stars (one in the Small Magellanic Cloud (SMC) and two in the Galaxy) that exhibited neither traces of carbon enhancements nor \sprocess elements, suggesting a likely failure of the third dredge-up (TDU). Further expanding on this, a recent investigation by \citet{Kamath2022} categorised single post-AGB stars with similar atmospheric parameters into two groups: those displaying \sprocess enrichment and those exhibiting no \sprocess enrichment.

Additionally, the majority of the post-AGB stars in binary systems (disc-sources) exhibit a chemical anomaly in their photosphere known as ‘chemical depletion’ \citep[see][and references therein]{oomen18,kamath19}. This phenomenon arises from the interaction between the binary post-AGB star and the surrounding circumbinary disc (disc-binary interaction). Their photospheric abundance pattern resembles the gas phase of the interstellar medium, with volatile elements like Zn and S maintaining their initial abundances, while refractory elements, including \sprocess elements, are notably underabundant, making them extrinsically metal-poor. 

In this paper, we focus on a subset of chemically peculiar post-AGB disc-sources that show carbon and \sprocess enrichment in their photospheric chemical composition, contrary to the commonly observed photospheric chemical depletion typically observed in disc-sources. We present a detailed atmospheric parameter and chemical abundance analysis of J005107.19-734133.3 (hereafter referred to as J005107), which represents the first known post-AGB disc-source in the SMC with an \sprocess enrichment. The \sprocess enrichment of this star was initially identified in the SMC survey by \citet{Kamath2014}. In addition, we incorporate two previously recognised \sprocess enriched post-AGB disc-sources: MACHO\,47.2496.8 in the Large Magellanic Cloud (LMC) \citep{Reyniers2007} and HD\,158616 in our Galaxy \citep{desmedt2016}. This provides an overview of carbon and \sprocess enriched binary post-AGB stars across both the Galaxy and the MCs. The unusual \sprocess enrichment in these stars suggests that the disc-binary interaction did not induce a photospheric chemical depletion. Through this study, we aim to understand the efficiency of chemical depletion in these systems and the effects of binarity on their peculiar chemical composition. We also investigate the underlying mechanism responsible for this chemical peculiarity.

This paper is structured as follows: Section~\ref{sec:target_data}, presents an overview of the targets in this study, providing detailed information on the photometric and spectroscopic data associated with each target. In Section~\ref{sec:spectral_analysis}, we provide a detailed description of our spectral analysis, to derive the atmospheric parameters and chemical abundances of the target stars. In Section~\ref{sec:phot_analysis}, we detail SED fitting, luminosity derivation and initial mass estimation of the targets. Finally, in Section~\ref{sec:discussion}, we investigate the chemical peculiarity of our targets and their underlying mechanism by comparing the observationally derived chemical abundances with theoretical predictions from the ATON evolutionary models \citep{ventura98}.

\section{Targets and Observations}
\label{sec:target_data}
\begin{figure}
    \includegraphics[width=\columnwidth]{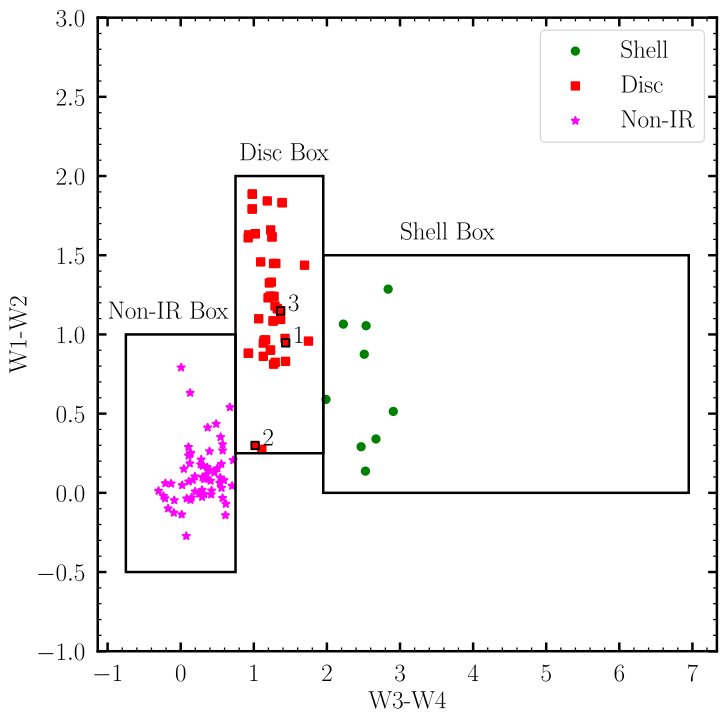}
    \caption{The WISE colour-colour diagram for the sample stars along with the stars studied in \citet{Gezer2015} is depicted here. The different types of SED characteristics among the post-AGB stars are represented here with different symbols and colours. The Non-IR box represents the non-dusty stars that show no IR excess in their SEDs. The sample stars are numbered according to their position in Table~\ref{tab:sample} for reference.}
    \label{fig:discShellPlot}
\end{figure}

The target sample for this study consists of three chemically peculiar objects previously classified as post-AGB stars, that have a ‘disc-type' SED (see Figure~\ref{fig:SED_4Sample}) and lie within the ‘disc box' of \citet{Gezer2015} (see Figure~\ref{fig:discShellPlot}). The three targets are J005107 from the SMC \citep{Kamath2014}, MACHO\,47.2496.8 from the LMC \citep{reyniers05}, and HD\,158616 \citep{desmedt2016} from the Galaxy. We note that HD 158616 is a confirmed binary according to orbital parameter studies conducted by \citet{oomen18}. With regards to J005107 and MACHO\,47.2496.8, taking into account a combination of their SED characteristics and pulsation features (as detailed below for each of the targets), we conclude that these two stars also reside in binary systems.

In Table~\ref{tab:sample}, we present our target sample including their other names and two sets of stellar parameters. The stellar parameters under "This Study" represent the spectroscopically derived atmospheric parameters in this study. The stellar parameters under "Literature" represent the spectroscopically determined atmospheric parameters from previous Literature analysis. 

\begin{sidewaystable}
    \captionsetup{justification=raggedright,singlelinecheck=false}
    \caption{Target sample along with two sets of stellar parameters. The final column lists the references to the previous literature analysis.}
        \begin{center}
            \begin{tabular}{ c l l c c c c c c c c c c  }
                \hline
                \addlinespace
            \#ID & Object Name & Other Name & \begin{tabular}[c]{@{}l@{}}\Teff\\ {(}K{)}\end{tabular} & \begin{tabular}[c]{@{}l@{}}\logg \\ {(}dex{)}\end{tabular} & \begin{tabular}[c]{@{}l@{}}\feh \\ {(}dex{)}\end{tabular} & \begin{tabular}[c]{@{}l@{}}\mv \\ {(}\kms{)}\end{tabular} & & \begin{tabular}[c]{@{}l@{}}\Teff \\ {(}K{)}\end{tabular} & \begin{tabular}[c]{@{}l@{}}\logg \\ {(}dex{)}\end{tabular} & \begin{tabular}[c]{@{}l@{}}\feh \\ {(}dex{)}\end{tabular} & \begin{tabular}[c]{@{}l@{}}\mv \\ {(}\kms{)}\end{tabular} & Ref. \\
            \addlinespace
                \hline
                \addlinespace
        		\multicolumn{3}{ c }{} & \multicolumn{4}{c }{This Study} & \multicolumn{5}{c }{Literature} \\
                \addlinespace
                \hline
                \addlinespace
                1 & J005107.19-734133.3 &	OGLE SMC-T2CEP-18	& 5768 $\pm$ 85	&	0.21 $\pm$ 0.10	&	-1.57 $\pm$ 0.10   &	4.08 $\pm$ 0.05	& & 5267 $\pm$ 250	&	0.72 $\pm$ 0.50	&	-1.56 $\pm$ 0.50  & ... & 1	\\
                2 & MACHO\,47.2496.8 &	OGLE LMC-T2CEP-15	&
                4898 $\pm$ 100	&	0.00 $\pm$ 0.50	&	-1.42 $\pm$ 0.16   &	4.00 $\pm$ 1.00	& & 4900 $\pm$ 250	&	0.00 $\pm$ 0.50	&	-1.50 $\pm$ 0.50   &	3.50 $\pm$ 1.00  &	2\\
                3 & HD\,158616 &	IRAS 17279-1119	& 7379 $\pm$ 110	&	1.47 $\pm$ 0.25	&	-0.50 $\pm$ 0.20   &	2.97 $\pm$ 0.25 &  & 7250 $\pm$ 125	&	1.25 $\pm$ 0.25	&	-0.64 $\pm$ 0.12 &	3.00 $\pm$ 0.50 &	3\\
                \addlinespace
                \hline
            \end{tabular}
        \end{center}
        \begin{tablenotes}
        
        \small
    \item \textbf{Notes:} The stellar parameters under "This study" represent the spectroscopically determined atmospheric parameters obtained using the $Python$ wrapper of iSpec (E-iSpec) (see section~\ref{sec:spectral_analysis}) whereas the stellar parameters under the title "Literature" represent the spectroscopically determined atmospheric parameters from the Literature analysis. Note that the low-resolution spectral analysis of J005107.19-734133.3 did not provide an estimated \mv\ value. Detailed information on the tabulated information can be found in the individual studies in column 'Ref'. The column 'Ref.' indicates the individual study: 1 - \cite{Kamath2014}, 2 - \cite{Reyniers2007}, 3 - \cite{desmedt2016}\\
        \end{tablenotes}
        \label{tab:sample}
\end{sidewaystable}

In the following paragraphs, we discuss the individual targets of our study in detail.\\

\begin{figure*}
    \begin{minipage}{0.5\textwidth}
        \includegraphics[width=\linewidth]{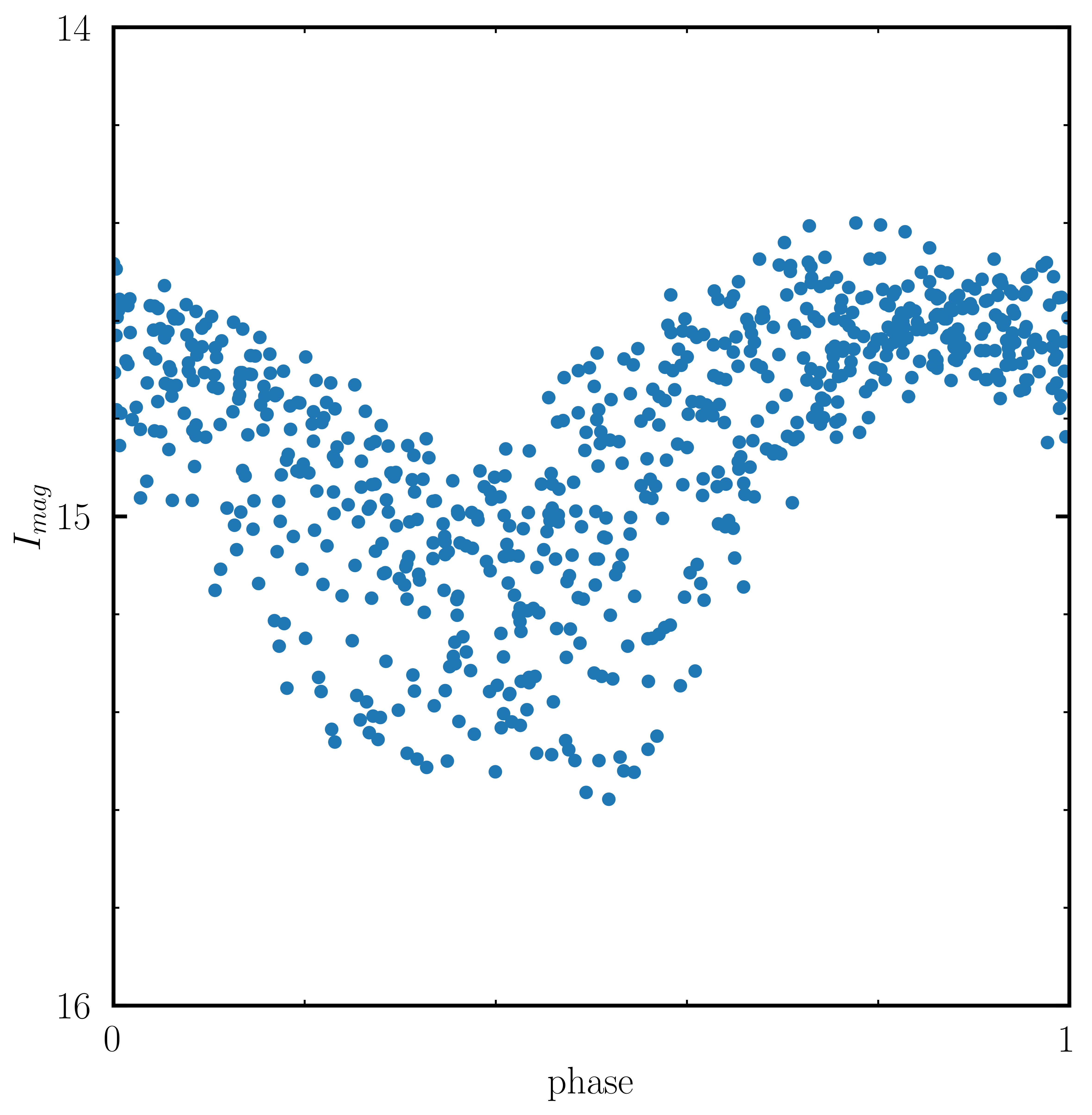}
        \label{subfig:lightCurve_J005107}
    \end{minipage}%
    \begin{minipage}{0.5\textwidth}
        \includegraphics[width=\linewidth]{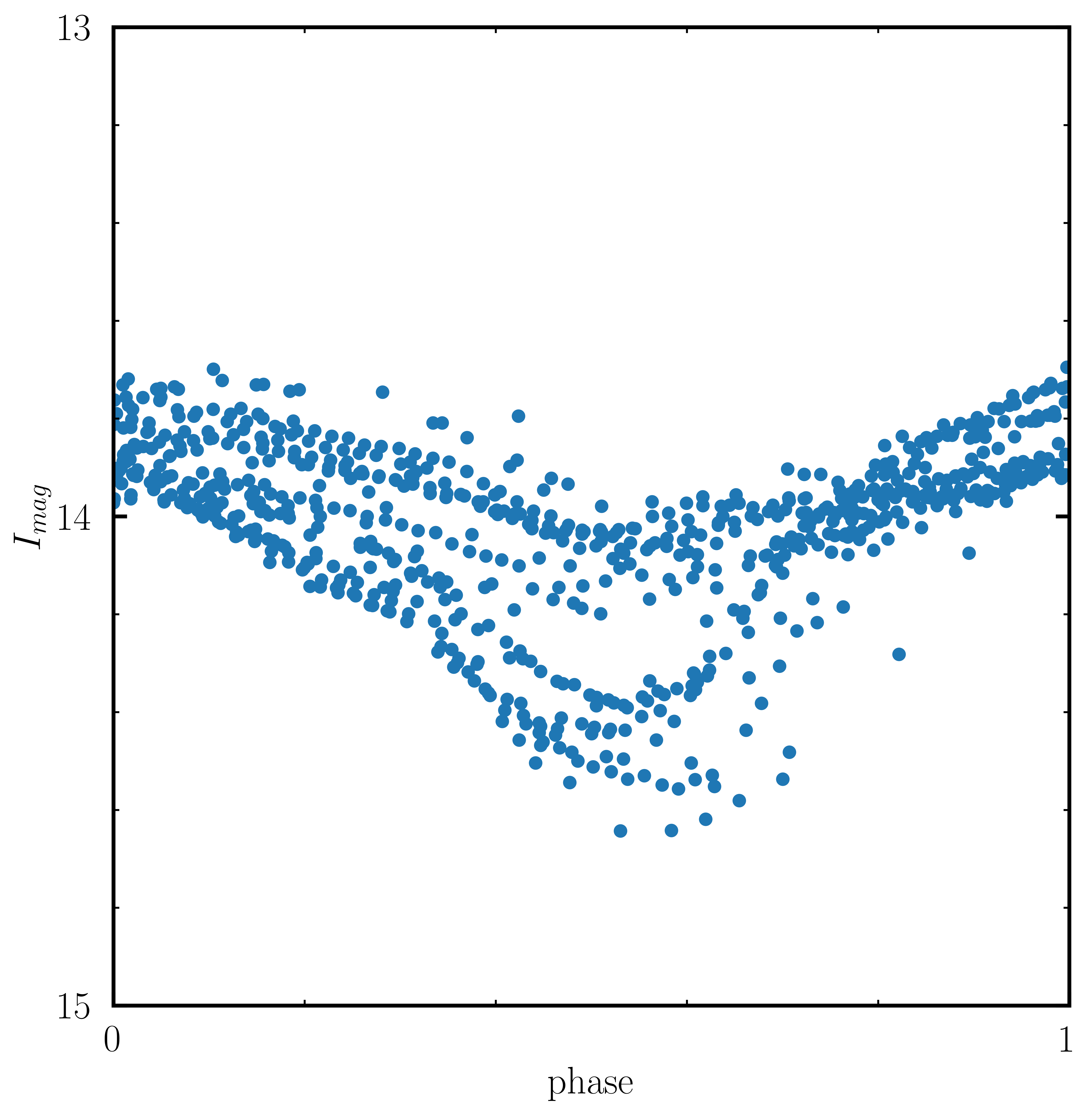}
        \label{subfig:lightCurve_macho47}
    \end{minipage}%
    \caption{The phased LCs of the RV Tauri targets J005107 (left) and MACHO\,47.2496.8 (right). The fundamental pulsation period used to phase the LC of each target is mentioned in the text. The LCs are scattered due to their semi-regular nature; they usually show considerable variations from cycle to cycle. This behaviour is typically more pronounced for the longer-period RV Tauri stars. The photometric data for J005107 and MACHO\,47.2496.8 have been sourced from the latest OGLE IV database of Type II Cepheids \citep{Soszy2018}.}
    \label{fig:light_Curve}
\end{figure*}

\textbf{J005107} has been classified as a post-AGB star and an RV Tauri pulsator in the photometric study conducted by \citet{Manick2018}. Additionally, \citet{Manick2018} categorised J005107 as an RVa type due to the lack of prominent long-term variability, based on its OGLE III light curve (LC). The fundamental period of this star is approximately 39.67 days according to the latest OGLE IV database, \citep{Soszy2018}. In Figure~\ref{fig:light_Curve} (left), we present the LC of J005107 from the latest data release: OGLE IV. Furthermore, \citet{Manick2018} also identified J005107 as a disc-source due to the presence of a characteristic IR excess in its SED (see Figure~\ref{fig:SED_4Sample}), consistent with the presence of a circumbinary disc, which is typical in binary post-AGB stars. Notably, the prominent Ba II features in the low-resolution optical AAOmega spectra presented in \citet{Kamath2014} recognised J005107 as the first \sprocess enriched RV Tauri star in the SMC.

\textbf{MACHO\,47.2496.8} was initially classified both as a post-AGB star and as an RV Tauri pulsator in the MACHO Project: LMC Variable Star Inventory VII, conducted by \citet{Alcock_1998} on optically bright post-AGB stars in the LMC. Additionally, \citet{Manick2018} classified MACHO\,47.2496.8 as an RVa type, based on its OGLE III LC. The fundamental period of this star is approximately 56.48 days according to the latest OGLE IV database \citep{Soszy2018}. In Figure~\ref{fig:light_Curve} (right), we present the LC of MACHO\,47.2496.8 from the latest data release: OGLE IV. \citet{Manick2018} also identified MACHO\,47.2496.8 as a disc-source due to the presence of a characteristic IR excess in its SED. In a series of low-resolution spectra described by \citet{Pollard_2000}, MACHO\,47.2496.8 exhibited strong C2 bands and Ba II features, leading to its identification as the first carbon and \sprocess enriched RV Tauri star. This classification was further confirmed by a subsequent high-resolution spectral analysis by \citet{Reyniers2007}. 

\textbf{HD\,158616} has been classified as an optically bright post-AGB star with a carbon enrichment, as determined by the spectroscopic study of \citet{Hans1997}. Additionally, the high-resolution spectroscopic studies by \citet{rao12} and \citet{desmedt2016} established HD\,158616 as the first post-AGB star rich in \sprocess elements discovered to be in a spectroscopic binary. HD\,158616 is confirmed to be a binary star with an orbital period of 365.0 days, and additional orbital properties can be found in \citet{oomen18}. \citet{desmedt2016} also identified this star as a disc-source due to a characteristic IR excess in its SED.

The SED fitting of the target stars is detailed in Section~\ref{sec:sed}.

\subsection{Photometric Data}
\label{sec:photometric_data}
To construct the SEDs in Section~\ref{sec:sed}, we collected the photometric data from the Vizier database \citep{Vizier2000}. Photometry at optical and near-IR wavelengths characterises the photospheric emission of the post-AGB star. Subsequently, the post-AGB photosphere is fitted using these data points (see Section~\ref{sec:sed}). The most significant photometry bands we used at these wavelengths are the UBVRI Johnson-Cousins bands \citep{Bessell1990}. For one of the targets, we also used photometry from Geneva \citep{Geneva1999} and SDSS \citep{SDSS2017}. Emission at longer wavelengths is dominated by low-temperature emission from the dusty disc. Therefore, photometry at mid- and far-IR wavelengths defines the circumstellar environment. To characterise this we use the photometric data from the 2MASS All-Sky Catalog of Point Sources \citep{Cutri2003}, the WISE All-Sky Data catalogue \citep{Cutri2012}, the AKARI/IRC mid-IR all-sky survey \citep{Ishihara2010}, and the IRAS catalogue of Point Sources \citep{IRAS}. \begin{sidewaystable}
    \captionsetup{justification=raggedright,singlelinecheck=false}
    \caption{Photometric data of the targets. See text for full details.}
    \resizebox{\linewidth}{!}{%
    \centering
    \begin{tabular}{ c c c c c c c c c c c c c c c c c c c c c }
        \hline
        \addlinespace
        Object Name & RA (deg) & Dec (deg) & U & B & SDSS.RP & V & R & I & J & H & K & W1 & W2 & AKARI.S9W & IRAS.F12 & W3 & W4 & IRAS.F25 & IRAS.F60 & IRAS.F100\\
        \addlinespace
        \hline
        \addlinespace
        J005107  & 12.780083 & -73.692598 & 16.65 & 16.657 & 99.99 & 15.84 & 15.45 & 14.85 & 14.33 & 13.65 & 12.93 & 11.39 & 10.44 & 99.99 & 99.99 & 8.365 & 6.926 & 99.99 & 99.99 & 99.99 \\
        MACHO 47.2496.8  & 73.930267 & -67.852718 & 17.31 & 16.23 & 99.99 & 15.08 & 14.28 & 14.00 & 13.19 & 12.67 & 12.57 & 12.09 & 11.79 & 99.99 & 99.99 & 10.26 & 9.248 & 99.99 & 99.99 & 99.99 \\
        HD 158616 & 262.695520 & -11.368968 & 11.92 & 10.09 & 9.368 & 9.642 & 99.99 & 99.99 & 7.845 & 7.299 & 6.525 & 5.178 & 4.029 & 3.301 & 3.520 & 2.600 & 1.234 & 2.900 & 1.600 & 1.980 \\
        \addlinespace
        \hline
    \end{tabular}
    }
    \begin{tablenotes}
        \small
        \item \textbf{Notes:} The RA and DEC coordinates are given for the J2000 epoch. Null magnitudes are listed as 99.999.
    \end{tablenotes}
    \label{tab:photometry}
\end{sidewaystable}The photometric magnitudes of all the objects covering the optical, near-IR, and mid-IR bands are presented in Table~\ref{tab:photometry}. As these stars are pulsating, multiple measurements for magnitude values within the same photometric bands are available. Hence, we provide the mean value of each measurement in Table~\ref{tab:photometry}. We refer to the Appendix A of \citet{oomen18} for the list of the most common catalogs used.

Figure~\ref{fig:SED_4Sample} shows the SEDs of all the targets in our sample. As mentioned in Section~\ref{sec:target_data}, the SEDs of our target stars are of ‘disc-type'.

\subsection{Spectroscopic Observations and Data Reduction}
\label{sec:spectro_data}
\begin{figure}
    \centering
    \includegraphics[width=.99\linewidth]{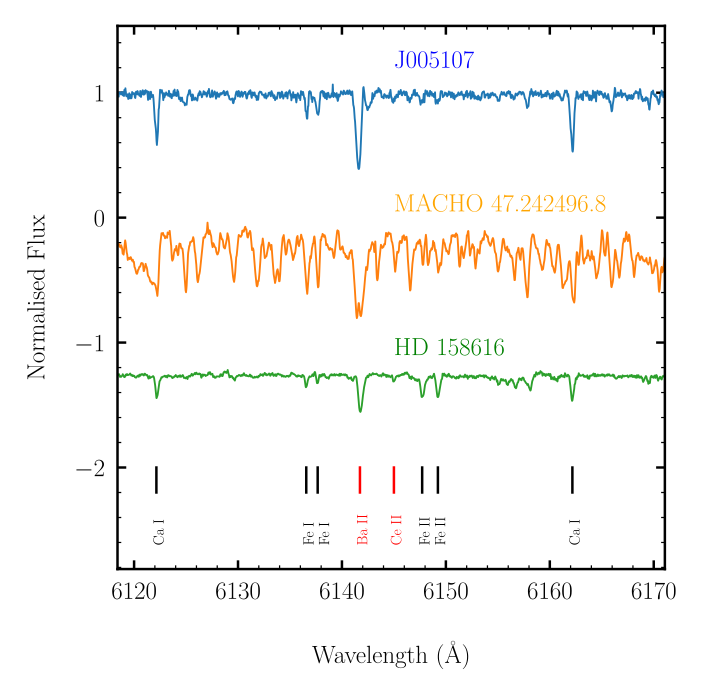}
    \includegraphics[width=.99\linewidth]{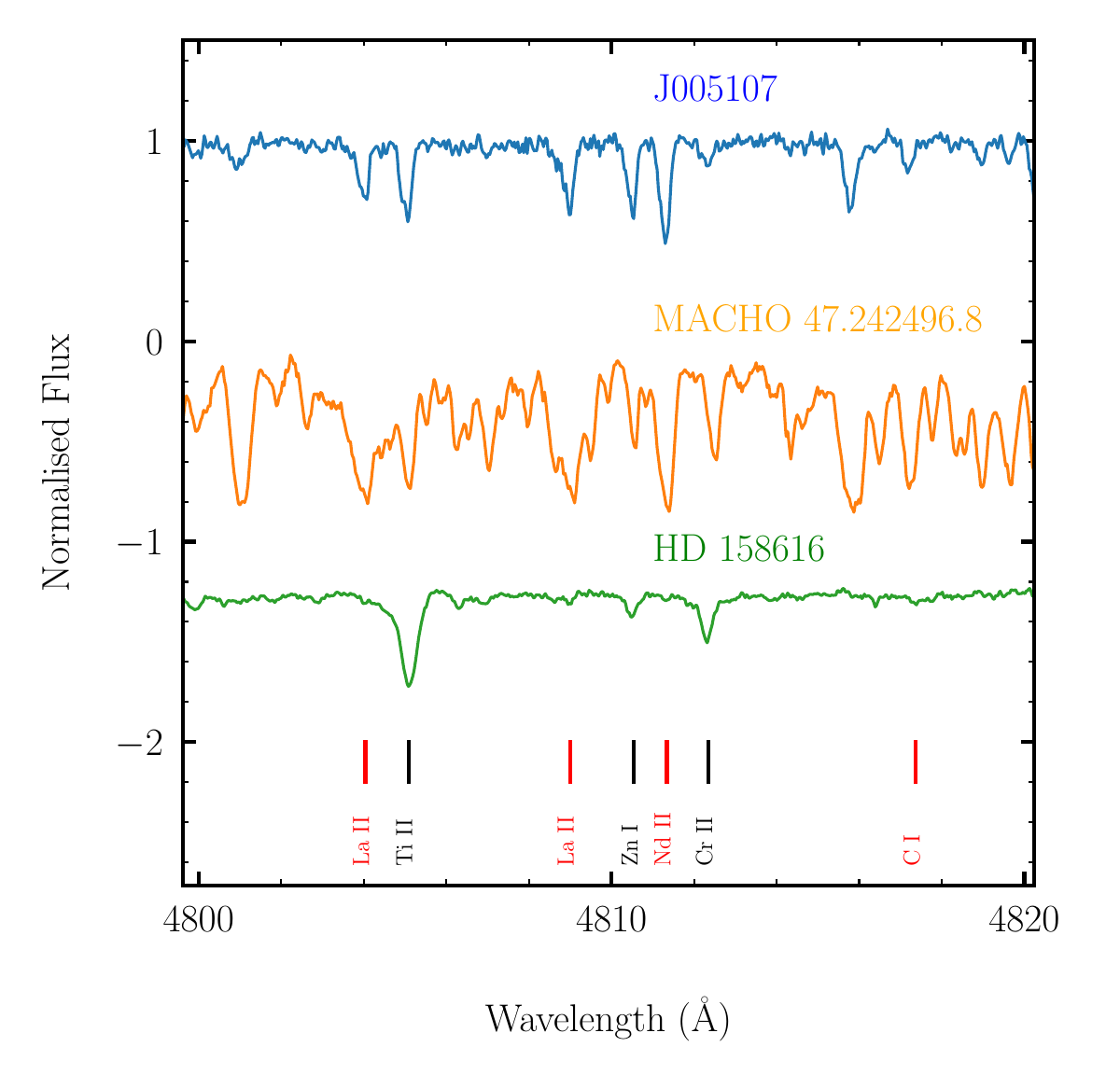}
    \caption{Comparison of the normalised and radial velocity corrected spectra of all target stars. The spectra have been shifted in flux for clarity. Red and black vertical lines mark the positions of \textit{s}-process elements (and also carbon) and non-\textit{s}-process elements, respectively. For more information, see the text.}
    \label{fig:Spectra}
\end{figure}

The high-resolution optical spectra of the three targets: J005107, MACHO\,47.2496.8 and HD\,158616 were taken from the Ultraviolet and Visual Echelle Spectrograph (UVES, \citet{UVES}), which is the echelle spectrograph mounted on the 8 m UT2 Kueyen Telescope of the Very Large Telescope (VLT) array at the Paranal Observatory of European Southern Observatory (ESO) in Chile, aiming to obtain comprehensive data on a large sample of post-AGB objects. The dichroic beam-splitter was used to get maximum wavelength coverage resulting in a wavelength coverage from approximately 3280 {\AA} to 4560 {\AA} for the blue arm of UVES, and from approximately 4726 {\AA} to 5800 {\AA} and 5817 {\AA} to 6810 {\AA} for the lower and upper part of the red arm of the UVES CCD chip respectively. There are small spectral gaps between 4560 {\AA} and 4726 {\AA} and between 5800 {\AA} and 5817 {\AA} due to the spatial gap between the three UVES CCDs. Each wavelength range is observed separately with a specific exposure time. The resolving power of the UVES spectra varies between $\sim\!60,000$ and $\sim\!65,000$. In Figure~\ref{fig:Spectra}, two spectral snippets are presented, illustrating the quality of the spectral data and highlighting the identification of certain \sprocess elements in the target stars.
\begin{table*}
    \caption{Observational logs of the target stars. The references for previous high-resolution spectroscopic studies utilising the spectra presented in this table are provided.}
    \begin{center}
            
        \begin{tabular}{ c c c c c c c c }
            \hline
            \addlinespace
        Name & \begin{tabular}[c]{@{}c@{}}
        Date \\ {(}YYYY-MM-DD{)}\end{tabular} & \begin{tabular}[c]{@{}c@{}}
        UTC Start \\ {(}hh:mm:ss{)}\end{tabular} &  \begin{tabular}[c]{@{}c@{}}Exp Time\\ {(}s{)}\end{tabular}&
        \begin{tabular}[c]{@{}c@{}}
        Telescope+ \\  Spectrograph
        \end{tabular} & \begin{tabular}[c]{@{}c@{}}
        Wavelength Coverage \\  {(}{\AA}{)}
        \end{tabular} & S/N & References \\
        \addlinespace
            \hline
            \addlinespace
            J005107.19-734133.3  &	2017-06-28	&	08:53:42	&	All: 3 × 3005$^{(1,2)}$   & VLT + UVES & 3281.91-6832.27 & 80 & - \\
            & 2017-07-16 & 08:09:24 & All: 3 × 3005$^{(3)}$ & VLT + UVES & 3281.91-6832.27 & 80 & \\
            \addlinespace
            \hline
            \addlinespace
            MACHO 47.2496.8 &	2005-02-09	&	00:30:00 & Blue: 7200   & VLT + UVES & 3758-4983 & 40 & a\\
            & 2005-02-08 & 02:49:00 & RedL: 7200 & VLT + UVES & 4780-6808 &  70 & \\
            & 2005-02-09 & 00:30:00 & RedU: 7200 & VLT + UVES & 6705-10084 &  80 & \\
            \addlinespace
            \hline
            \addlinespace
            HD 158616  & 2014-09-18	& 00:00:00 & All: 1 × 65   & VLT + UVES & 3280-6810 & 80 & b\\
            \addlinespace
            \hline
        \end{tabular}
    \end{center}
    \begin{tablenotes}
     \small
    \item \textbf{Notes:} The exposure times of UVES spectra are split into four categories: exposure times for the Blue arm, Red Lower arm, Red Upper arm or all three arms; the latter is indicated with “All”. The wavelength coverage for the "Blue" UVES arm is from approximately 3280 to 4530 {\AA} and the "RedL" and "RedU" UVES arms are from approximately 4780 to 5770 {\AA} and from 5800 to 6810 {\AA}, respectively.\\
    $^{(1,2)}$ The log details are for Observation one (OB1) and Observation two (OB2) together. (see text for details on OB1, OB2 and OB3).\\
    $^{(3)}$ The log details are for Observation three (OB3).\\
    \item \textbf{References:} a - \cite{Reyniers2007}, b - \cite{desmedt2016}\\ 
    \end{tablenotes}
    \label{tab:log}

\end{table*}

Table~\ref{tab:log} gives the log of the observations, the spectral intervals covered and the signal-to-noise (S/N) ratio for each object. In general, the signal-to-noise (S/N) ratio is lower at blue wavelengths. We note that the echelle data does not have a consistent S/N because it depends on the blaze function. The S/N is higher at the blaze wavelengths, where the spectrograph works most efficiently, compared to other parts of the spectrum. The "References" column points to the literature of the previous high-resolution studies.

In the following paragraphs, we only discuss the spectral reduction, normalisation, radial velocity correction and weighted average merging of J005107 in detail. Similar steps of the other two targets: MACHO\,47.2496.8 and HD\,158616 are explained in detail by \citet{Reyniers2007} and \citet{desmedt2016}, respectively.

The UVES spectral reduction of the raw data of J005107 was performed using the UVES reduction pipeline from EsoReflex on the default optimal settings. The EsoReflex software allows the user to choose from a variety of pipelines that correspond to the various telescopes available at ESO. The reduction process involves the standard steps of extracting frames, determining wavelength calibration and applying this scale to flat-field divided data. As part of the reduction, cosmic clipping was also taken into account.

The normalisation of the reduced spectrum was done by fitting in small spectral windows with fifth-order polynomials through interactively defined continuum points. We note that normalisation is the most delicate step in the reduction procedure, especially in the blue part of the spectrum, where the spectrum is so crowded that the continuum is rarely reached. The most significant source of uncertainty for the abundances calculated from lines in this region is the continuum location. Three observations were taken for J005107 with an exposure time of 3005 s for each UVES arm. Since the third observation (OB3) of J005107 was taken after 19 days (see Table~\ref{tab:log}) from observation one (OB1) and observation two (OB2), and considering the target to be an RV Tauri pulsator, the phase of OB3 is different from that of OB1 and OB2. Hence, we decided to treat the OB3 as an independent spectrum for the rest of our analysis. However, the OB3 spectra had minimal lines to be useful for both atmospheric parameter analysis as well as abundance derivation. Therefore, we chose to use only OB1 and OB2 for the rest of our analysis.

To determine the radial velocity of J005107, several well-identified atomic lines in the spectrum were fitted with a Gaussian curve to find their centre wavelength. The Doppler shift equation was used to compute the shifted velocity. This yields a heliocentric radial velocity of $v=160\pm$8 \kms\ for OB1 and OB2, which is precise enough for line identification purposes. This is also in good agreement with the radial velocity value $v=187.5\pm4.3$ \kms\ estimated by \citet{Kamath2014} using the low-resolution spectra. Furthermore, the velocity of J005107 validates the membership of the SMC, with the heliocentric radial velocity of the SMC being $\sim\!160$ \kms\ \citep{Richter1987}.

Once all three spectra (Blue, RedL and RedU) of OB1 and OB2 were normalised and radial velocity corrected, a weighted mean average merging was performed, thereby obtaining a single final spectrum, which was used to perform a detailed spectral analyses of J005107. A significant portion of the Blue spectrum (wavelength range $3280-4560$ {\AA}) has a signal-to-noise ratio (S/N) that is too low to be useful for a precise spectral abundance analysis; as a result, these wavelength ranges are not used for the spectral analyses of J005107.

\section{Spectroscopic Analysis}
\label{sec:spectral_analysis}
We performed a systematic spectral analysis for all our target stars, which included: (a) precise estimation of atmospheric parameters: effective temperature (\Teff), surface gravity (\logg), metallicity (\feh), and microturbulant velocity (\mv) and (b) chemical abundance derivation for all identifiable elements from the final prepared spectrum of each star. The atmospheric parameter determination, as well as the abundance analysis, are carried out using E-iSpec, our own $Python-$based semi-automated spectral analysis tool for optical and NIR spectra, serving as a wrapper for iSpec \citep{Blanco2014a,Blanco2019}. E-iSpec offers enhanced capabilities for determining atmospheric parameters, elemental abundances, and isotopic ratios in evolved stars with complex atmospheres, utilising 1D local thermal equilibrium (LTE) model atmospheres, MOOG radiative transfer code \citep{sneden1973}, and comprehensive line lists. A detailed description of E-iSpec and our spectral analysis procedure is presented in Mohorian et al. 2024.

In brief, there are two modules in E-iSpec: one for determining atmospheric parameters (using the equivalent width (EW) method, see Section~\ref{sec:atmos_param}) and the other for determining abundances (using EW and synthetic spectral fitting (SSF), see Section~\ref{sec:abund_analysis}). For the spectral analyses, the latest LTE model atmospheres, ATLAS9 \citep{castelli03} or MARCS \citep{MarcsModel2008}, chosen depending on the \Teff\ of the star, were used in combination with the LTE abundance calculation routine MOOG (version July 2009) by \citet{sneden1973}. The model atmospheres were uniquely determined by the \Teff, \logg, \feh\ and \mv. Line lists from the Vienna Atomic Line Database (VALD) \citep{VALDLinelist} with a range of 3000 {\AA} to 11000 {\AA} were used to identify the spectral lines of the target stars. This covers the full wavelength coverage of UVES spectra and allows for identifying spectral lines of about 160 ions ranging from He (Z = 2) up to U (Z = 92).

We note that the atmospheric parameters and chemical abundances of all our targets were derived aiming at isolated, unblended and non-saturated lines, the identification of which was challenging due to the small-amplitude pulsations and pulsation-driven shocks commonly observed in RV Tauri pulsating stars \citep{Reyniers2007}. We avoided spectral noise by using lines with EW greater than 5 m{\AA}. Additionally, we avoided lines with EW greater than 150 m{\AA} since they are saturated. We note that non-LTE effects were not taken into account for our analysis. 

In the following subsections, we briefly describe our spectral analyses: atmospheric parameter determination (see Section~\ref{sec:atmos_param}) and derivation of chemical abundances (see Section~\ref{sec:abund_analysis}). 

\subsection{Atmospheric parameter determination}
\label{sec:atmos_param}
We derived the atmospheric parameters of all the target stars using Fe I and Fe II lines (see Section~\ref{sec:spectral_analysis} for more details on line selection). Fe was chosen due to its large number of unblended lines, covering a broad range of excitation potential. The EWs were calculated for each line using an iterative procedure in which the theoretical EWs of individual lines were compared to the calculated EWs. The atmospheric parameter determination method is outlined below. 

The effective temperature (\Teff) was estimated using excitation analysis, wherein the abundances derived from Fe I lines are imposed to be independent of their excitation potential (EP). An accurate derivation of \Teff\ is promoted by the wide range in EP (0.5 - 5 eV) and the sufficiently large number of lines. The surface gravity (log g) is determined using ionisation analyses, imposing the abundances to be independent of the lines being from neutral or ionised Fe. The microturbulent velocity (\mv) was derived by imposing the independence of the abundance derived from individual Fe lines of the reduced equivalent width (RW). Concerning the error estimation for atmospheric parameters, they are determined from the covariance matrix generated by the non-linear least-squares fitting algorithm in E-iSpec. 

We note that the atmospheric parameter analysis can be verified using other species like Ti or Cr, provided there is a substantial number of lines for certain species. However, due to a large number of blends, the majority of the elements only have less than three useful lines to perform the analysis. 

The results of the atmospheric parameter analysis of our targets are presented in Table~\ref{tab:sample} and discussed in Section~\ref{sec:results_J005107}.

\subsection{Abundance Analysis}
\label{sec:abund_analysis}
We derived chemical abundances using both the EW method and the SSF technique - which are modules of E-iSpec. In the EW method, observed EWs were compared with predictions from theoretical model atmospheres (ATLAS9 \citep{castelli03} or MARCS \citep{MarcsModel2008}), involving an iterative adjustment of assumed abundances until observed and predicted EWs converged, yielding best-fit abundance values. The SSF technique involved comparing observed stellar spectra with synthetic spectra generated by theoretical model atmospheres, employing chi-square fitting to iteratively refine model parameters and derive precise chemical abundances.

We note that the lines used for the abundance derivation are as per the criteria mentioned in Section~\ref{sec:spectral_analysis}. The complete linelists of the target stars are provided as online supporting material.

The uncertainties associated with the derived abundances are determined using the procedure outlined by \citet{Deero2005}. E-iSpec provides the errors of the derived abundances only as standard deviations of the measured values (\sigmalinetoline). However, the total error \sigmatotal\ is the quadratic sum of \sigmalinetoline, uncertainties in abundances due to atmospheric parameters and the \feh\ error. To determine the sensitivity of the abundances to the stellar parameters, the abundances were recalculated changing \Teff, \logg\ and \mv\ with their calculated error (see Table~\ref{tab:sample} for atmospheric parameter errors). We impose a \sigmalinetoline\ of 0.20 dex for all the ions for which only one line was available. The chosen uncertainty of 0.2 dex corresponds to the allowed standard deviation for the line-to-line scatter, indicating the range within which all individual line abundances are expected to lie.

The results of the abundance analysis of our targets are presented in Table~\ref{tab:abund_J005107} and Table~\ref{tab:appendix:abund_MAcho47&HD158616} and discussed in Section~\ref{sec:results_J005107}.

\subsection{Results of Spectroscopic Analysis: Atmospheric Parameters and Chemical Abundances}
\label{sec:results_J005107}
In this section, we present the results of the spectral analyses of J005107. We note that this study marks the first high-resolution spectral analyses of J005107. However, for MACHO\,47.2496.8 and HD\,158616 similar high-resolution spectral analyses have been previously carried out by \citet{Reyniers2007} and \citet{desmedt2016}, respectively. For benchmarking our methodology (i.e., the E-iSpec Code) we repeated the spectral analyses for these two objects (see \ref{sec:appendix:analysis2targets} for the final results). Since the derived atmospheric parameters and abundances of MACHO\,47.2496.8 and HD\,158616 align closely with values reported in the literature (refer to Table~\ref{tab:sample} and Table~\ref{tab:appendix:abund_MAcho47&HD158616}), we opted to adopt the literature values of both atmospheric parameters and abundances for the rest of our analysis.

The various atmospheric parameters of J005107 are displayed in Table~\ref{tab:sample}, under the title "This study". The atmospheric parameters of J005107 clearly fall within the range of typical post-AGB parameter values. 

\begin{table}
    \centering
    \caption{Spectroscopically determined abundance results for J005107.}
        \label{tab:abund_J005107}
        \begin{tabular}{ c c c c c c c c }
            \hline
            \addlinespace
            Ion & Z & \logepsilonsun & N & \xfe & \sigmatotal & \logepsilon & \sigmalinetoline \\
            \addlinespace
            \hline
            \addlinespace
            C I  & 6 & 8.43 & 1 & 1.09 & 0.217 & 7.95 & 0.20\\
            O I  & 8 & 8.69 & 1 & 1.06 & 0.212 & 8.18 & 0.20\\
            Mg I  & 12 & 7.60 & 2 & 0.77 & 0.160 & 6.79 & 0.15\\
            Si I  & 14 & 7.51 & 3 & 0.61 & 0.081 & 6.54 & 0.07\\
            Ca I  & 20 & 6.37 & 3 & 0.43 & 0.114 & 5.19 & 0.04\\
            Sc II  & 21 & 3.15 & 1 & 0.33 & 0.215 & 1.91 & 0.20\\
            Ti I  & 22 & 4.95 & 2 & 0.59 & 0.205 & 4.25 & 0.18\\
            Cr I  & 24 & 5.64 & 2 & 0.43 & 0.197 & 4.49 & 0.19\\
            Fe I  & 26 & 7.50 & 37 & 0 & 0.132 & 5.92 & 0.09\\
            Fe II  & 26 & 7.50 & 6 & 0 & 0.171 & 5.92 & 0.12\\
            Ni I  & 28 & 6.22 & 1 & 0.53 & 0.241 & 5.17 & 0.20\\
            Zn I  & 30 & 4.56 & 1 & 0.48 & 0.246 & 3.46 & 0.20\\
            Y II  & 39 & 2.12 & 5 & 1.13 & 0.115 & 1.76 & 0.03\\
            Zr II  & 40 & 2.58 & 2 & 0.74 & 0.101 & 1.74 & 0.02\\
            La II  & 57 & 1.10 & 3 & 1.52 & 0.133 & 1.04 & 0.04\\
            Ce II  & 58 & 1.58 & 7 & 1.44 & 0.138 & 1.44 & 0.04\\
            Pr II  & 59 & 0.72 & 3 & 1.66 & 0.211 & 0.8 & 0.20\\
            Nd II  & 60 & 1.42 & 15 & 1.69 & 0.146 & 1.53 & 0.07\\
            Sm II  & 62 & 0.96 & 4 & 1.70 & 0.147 & 1.08 & 0.07\\
            Gd II  & 64 & 1.07 & 2 & 1.66 & 0.113 & 1.15 & 0.05\\
            \addlinespace
            \hline
        \end{tabular}
    \begin{center}
    \begin{tablenotes}
     \small
    \item \textbf{Notes:} The ions that were detected and their corresponding atomic number (Z) are listed in columns 1 and 2, respectively. The solar abundances (\logepsilonsun) in column 3 are retrieved
from \citet{Asplund2009}. N represents the number of lines used for each ion, \xfe\ is the element-over-iron ratio, \sigmatotal\ is the total uncertainty on \xfe, \logepsilon\ is the derived abundance, and \sigmalinetoline\ is the line-to-line scatter. We impose a \sigmalinetoline\ of 0.20 dex for all ions for which only one line is available for the abundance determination.
    \end{tablenotes}
    \end{center}  
\end{table}

Table~\ref{tab:abund_J005107} presents the final derived abundances for different elements of J005107. Unfortunately, all Ba lines were severely saturated (see Figure~\ref{fig:Spectra} at 6141.7324 \AA), making it impossible to accurately determine the abundance of one of the significant \sprocess species. Except for the species where all observable lines turned out to be blended, the other \sprocess elemental abundances were derived from isolated single lines. We list the elements by their atomic number (Z), the solar abundances (\logepsilonsun) retrieved from \citet{Asplund2009}, the number of lines identified for each element (N), the element-over-iron ratio (\xfe), the total uncertainty on \xfe\ (\sigmatotal), the determined abundance (\logepsilon), and the line-to-line scatter (\sigmalinetoline). Although the abundances of several useful elements for nucleosynthesis studies could not be determined, Table~\ref{tab:abund_J005107} still provides quantified abundances of a range of \sprocess elements. See Section~\ref{sec:spectral_analysis} for details on atmospheric parameters and derived abundance uncertainties.

\begin{figure}
    \includegraphics[width=\columnwidth]{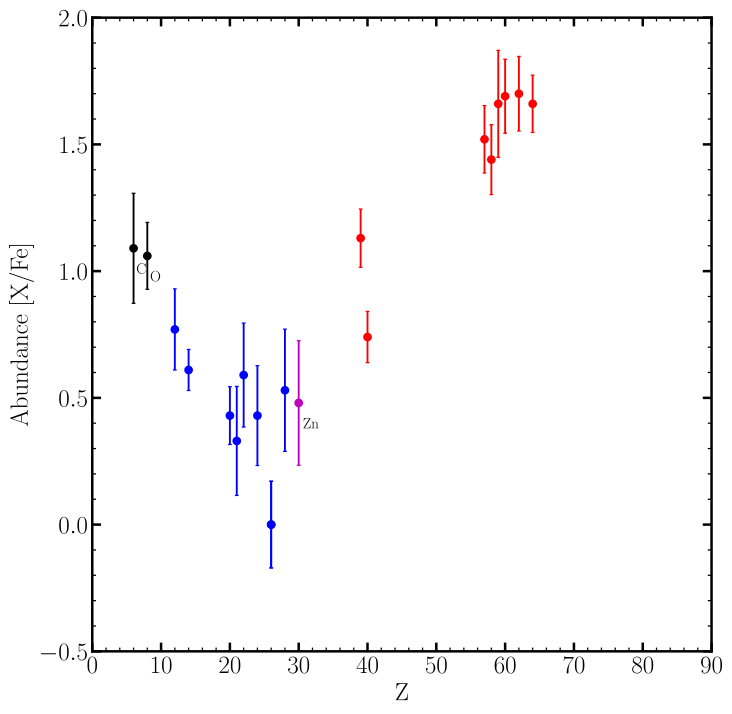}
    \caption{Spectroscopically derived abundances of J005107. The error bars represent the total uncertainties \sigmatotal. Some elements are labelled for reference. The black colour data points represent CNO elements, the blue represents Fe peak elements, the magenta represents Zn and S, and the red represents \sprocess elements.}
    \label{fig:abund_J005107}
\end{figure}
Combining the results of both atmospheric parameter analysis and abundance derivation we interpret the characteristics of J005107. J005107 is an F-type star (\Teff\ = $5764\pm85$ K) with a low surface gravity (\logg\ = $0.21\pm0.10$ dex) and an iron abundance (\feh\ = $-1.57\pm0.10$). The iron abundance of J005107 is low compared to the mean metallicity of the SMC \feh\ = -0.7 dex \citep{luck1998}. This categorises J005107 as a low-metallicity star, which is commonly recognised as an astrophysical production site producing heavy \sprocess elements provided the TDU occurs \citep{bisterzo2010}. Figure~\ref{fig:abund_J005107} shows the abundance distribution (\xfe) for J005107 as a function of atomic number (Z). From the abundance plot, it is clear that J005107 is strongly enriched in \sprocess elements with an \sfe\ = $1.52\pm0.20$ dex. Moreover, it also has a carbon enrichment with \cfe\ = $1.09\pm0.21$ dex. However, we note that owing to the observed blending in the single available carbon line (located at 4817.373 \AA\ as detailed in the linelist), accurately determining the carbon abundance posed a significant challenge. 

\section{Photometric Analysis}
\label{sec:phot_analysis}
In this section, we present details on the SED fitting (Section~\ref{sec:sed}) and the luminosity derivation (Section~\ref{sec:luminosity}) of our target stars.
\subsection{SED Fitting}
\label{sec:sed}
Most of our targets are RV Tauri pulsators that often show large amplitude pulsations that can go up to several magnitudes in V. One of the greatest challenges in fitting the SEDs of these stars is that these pulsations cause a scatter in the photometric data points. Therefore reddening is difficult to determine. To overcome this issue, it is crucial to have full coverage of all photometric bands in the same pulsation phase. However, since the same pulsation phase data is unavailable, we have utilised all the accessible photometric data points (see Section~\ref{sec:photometric_data}) to construct the SEDs. We determined the SED of all the targets of this study in a strictly homogeneous way, as explained below.

\begin{figure}
        \includegraphics[width=\columnwidth]{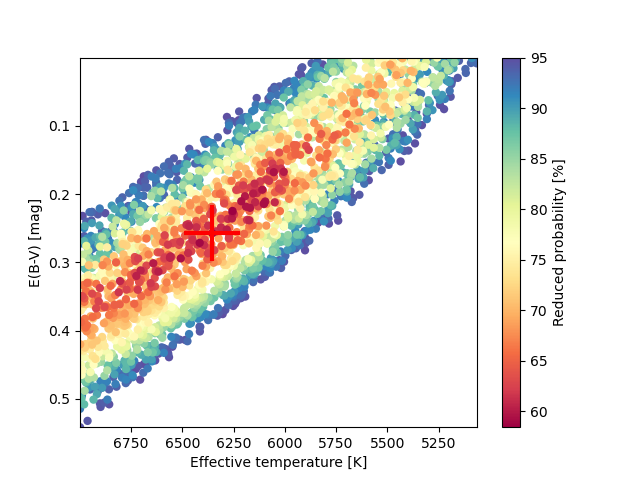}
    \caption{An example of $\chi^2$ plot of J005107 to obtain the reddening parameter $E(B-V)$ after the parameter grid search. This plot illustrates the correlation between \Teff\ and $E(B-V)$ of the targets.}
    \label{fig:TeffBC_4Sample}
\end{figure}

\begin{figure*}
    \begin{minipage}{0.5\textwidth}
        \includegraphics[width=\linewidth]{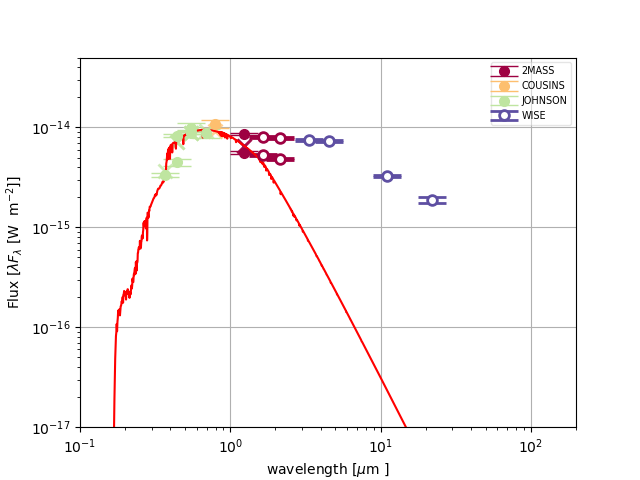}
        \label{subfig:sed_J005107}
    \end{minipage}%
    \begin{minipage}{0.5\textwidth}
        \includegraphics[width=\linewidth]{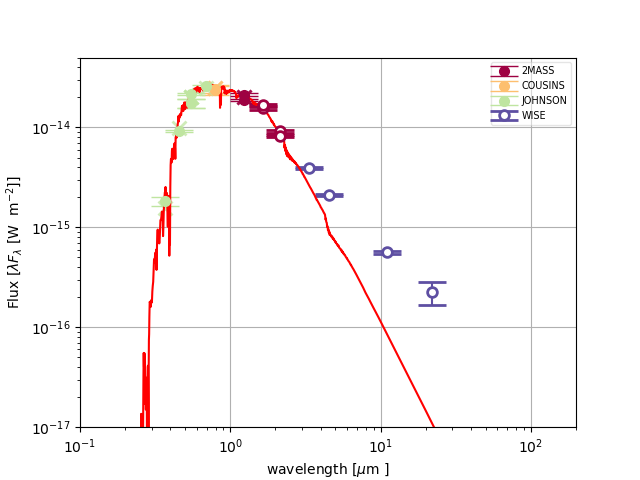}
        \label{subfig:sed_mach047}
    \end{minipage}\\[1ex]
    \begin{minipage}{0.5\textwidth}
        \includegraphics[width=\linewidth]{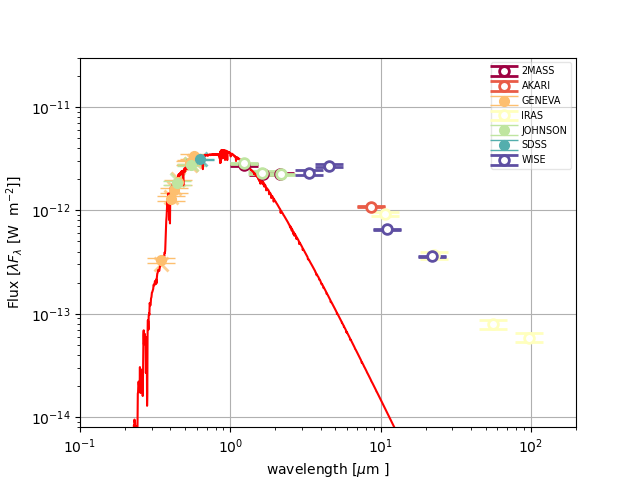}
        \label{subfig:sed_HD158616}
    \end{minipage}%
    \caption{SEDs of J005107 (top left), MACHO\,47.2496.8 (top right) and HD\,158616 (bottom). The data points represent the dereddened photometry. The red line represents the best-fitting scaled model atmosphere (see text for details).}
    \label{fig:SED_4Sample}
\end{figure*}

We adopted the same approach described in \citet{Kluska2022} and \citet{Kamath2022} to construct all the SEDs in this study. In summary, we began by determining the total line-of-sight reddening or extinction parameter ($E(B-V)$) by minimising the difference between the optical fluxes and the reddened photospheric models. The total reddening includes both interstellar and circumstellar reddening. We assume that the total reddening in the line of sight has the wavelength dependency of the interstellar-medium extinction law \citep{cardelli89} with Rv = 3.1. The extinction law in the circumstellar environment likely differs from the interstellar extinction law. However, investigating this aspect falls outside the scope of our current study. For the atmospheric models, we used the appropriate Kurucz atmospheric models \citep{castelli03}, the parameters of which were taken from the spectroscopic analysis presented in Section~\ref{sec:atmos_param} (see Table~\ref{tab:sample}). We interpolated in the $\chi^2$ landscape between the models centred around the spectroscopically determined parameters and applied ranges of $\Delta$\Teff\ $\pm 500$ K, $\Delta$\logg\ $\pm 0.5$ dex, $\Delta$\feh\ $\pm 0.5$ dex (see Figure~\ref{fig:TeffBC_4Sample}). The SEDs of our three target stars are presented in Figure~\ref{fig:SED_4Sample}.

\begin{table*}
    \centering
    \caption{Luminosities of the targets derived using the SED fitting method and PLC relation method.}
        \label{tab:Lumiosity_sample}
        \begin{tabular}{ c c c c c c c c c }
            \hline
            \addlinespace
            Object & \begin{tabular}[c]{@{}l@{}}$P_0$\\ {(}days{)}\end{tabular} & \begin{tabular}[c]{@{}l@{}}\Lsed\\ {(}\Lsun{)}\end{tabular} & \begin{tabular}[c]{@{}l@{}}$\Delta$\Lsed\\ {(}\Lsun{)}\end{tabular} & \begin{tabular}[c]{@{}l@{}}\Lplc\\ {(}\Lsun{)}\end{tabular} & \begin{tabular}[c]{@{}l@{}}$\Delta$\Lplc\\ {(}\Lsun{)}\end{tabular} & \begin{tabular}[c]{@{}l@{}}$E(B-V)$\\ {(}mag{)}\end{tabular} & SED type & Comments\\
            \addlinespace
            \hline
            \addlinespace
            J005107.19-734133.3 & 39.67 & 2702 & $^{+5108}_{-1747}$  & 2868 & $^{+6449}_{-1405}$ & $0.22^{+0.47}_{-0}$ & Disc & post-AGB, RV Tauri\\\\
            MACHO 47.2496.8 & 56.48 & 3001 & $^{+4051}_{-2775}$ & 2208 & $^{+4087}_{-1877}$ & $0.05^{+0.24}_{-0}$ & Disc & post-AGB, RV Tauri\\\\
            HD 158616 & - & 8256 & $^{+10301}_{-6288}$ & - & - & $0.53^{+0.63}_{-0.48}$ & Disc & post-AGB, Binary, \Porb=363.3 days \\\\
            \addlinespace
            \hline
        \end{tabular}
    \begin{center}
    \begin{tablenotes}
     \small
    \item \textbf{Notes:} $P_0$ is the fundamental pulsation period of the RV Tauri targets and are shown in Col. 2. The luminosities derived from SED along with their corresponding upper and lower limits are displayed in Col. 3 and 4 respectively. Similarly, the luminosities obtained using the PLC relation along with their corresponding upper and lower limits are displayed in Col. 5 and 6 respectively. Note that the target HD 158616 does not have luminosity derived from the PLC relation as it is not an RV Tauri star. The reddening derived from the SED model along with their corresponding upper and lower limits are shown in Col. 7, and the SED type of the targets are presented in Col. 8 (see text for details).
    \end{tablenotes}
    \end{center}  
\end{table*}

\subsection{Luminosity Determination}
\label{sec:luminosity}
Accurately determining the luminosities of post-AGB stars and combining them with their effective temperatures (\Teff) allows us to investigate their positions in the Hertzsprung-Russell (HR) diagram. This helps us better understand the evolutionary stage of the target post-AGB objects. However, accurately deriving the luminosities of post-AGB stars remains a significant challenge due to multiple factors. These include the impact of reddening—both circumstellar and interstellar—on observational photometric data, along with complexities arising from variability, particularly for the RV Tauri pulsators. Moreover, concerning binary post-AGB stars with orbital periods ranging from 100 to 700 days, systematic errors in parallax determination affect distances and consequently, luminosities \citep{Kamath2022}. This issue arises because Gaia EDR3 does not account for the orbital motion of binaries, often resulting in an underestimation of parallax. Consequently, this oversight impacts the derived distances and the luminosities of these stars \citep{Pourbaix2019}.

To estimate the luminosity of our targets, it is imperative to address the limitations above. We, therefore, use two methods discussed in the following subsections, i.e., the SED fit (\Lsed) (Section~\ref{sec:luminosity_sed}) and the period-luminosity-colour (PLC) relation (\Lplc) (Section~\ref{sec:luminosity_plc}). We find \Lplc\ to offer greater precision and reliability when compared to \Lsed. This preference arises due to significant variations in \Teff\ across the pulsation cycle in our target stars, which are RV Tauri pulsators (see Section~\ref{sec:target_data}). These fluctuations also lead to notable changes in the model fitting of stellar atmospheres (illustrated by the red lines in Figure~\ref{fig:SED_4Sample}). Additionally, as mentioned above, for stars in binary systems, the uncertainties in distances obtained from parallax measurements are markedly influenced by binary orbital motion, which adds to the uncertainties in \Lsed. 

\subsubsection{Deriving the Luminosity using SED}
\label{sec:luminosity_sed}
The bolometric luminosities (\Lsed) of the targets were derived by integrating the flux under the dereddened photospheric SED model (see Section~\ref{sec:sed}). To achieve this, we assumed an average distance of $62.1 \pm 1.9$ kpc to the SMC \citep{graczyk} and $49.97 \pm 1$ kpc to the LMC \citep{Walker2012,Pietrzy2013} respectively. For our Galactic target HD\,158616, we utilised the more precise Bailer-Jones geometric distances (i.e., $z_{\rm BJ}$) and their associated upper and lower limits ($z_{\rm BJU}$ and $z_{\rm BJL}$ respectively) from the study by \citet{BailerJones2021A}. The geometric distances were determined using Gaia EDR3 parallaxes, incorporating a direction-dependent prior distance. Throughout this computation, we assumed that the flux emitted by the stars is radiated isotropically. The derived SED luminosities (\Lsed) along with their corresponding upper and lower limits ($\Delta$\Lsed) for the targets are presented in Col. 3 and 4 of Table~\ref{tab:Lumiosity_sample}, respectively. To calculate these upper and lower limits for the SED luminosities ($\Delta$\Lsed), we utilised the corresponding upper and lower limits of the reddening values, as derived and outlined in Col 7 of Table~\ref{tab:Lumiosity_sample}.

\subsubsection{Deriving the Luminosity using PLC Relation}
\label{sec:luminosity_plc}
One of the greatest benefits of type II Cepheids is that their pulsations can be used to determine their luminosities due to the correlation between their pulsational periods and luminosities. Unlike the previous SED fitting method (see Section~\ref{sec:luminosity_sed}), this approach does not rely on distance dependencies, providing a valuable advantage for accurate luminosity derivation.

Many post-AGB stars, specifically RV Tauri stars exhibit strong radial pulsations because they are located in the long-period segment of the type II Cepheid instability strip. The radial modes of these stars can be directly linked to their physical sizes. By incorporating a colour or temperature factor, one can establish a connection between the star's luminosity and its pulsation period. This relationship is referred to as the period-luminosity-colour (PLC) relation, as shown in studies by \citet{Alcock_1998} and \citet{Ripepi2015}.

\begin{figure*}
    \begin{minipage}{0.5\textwidth}
        \includegraphics[width=\linewidth]{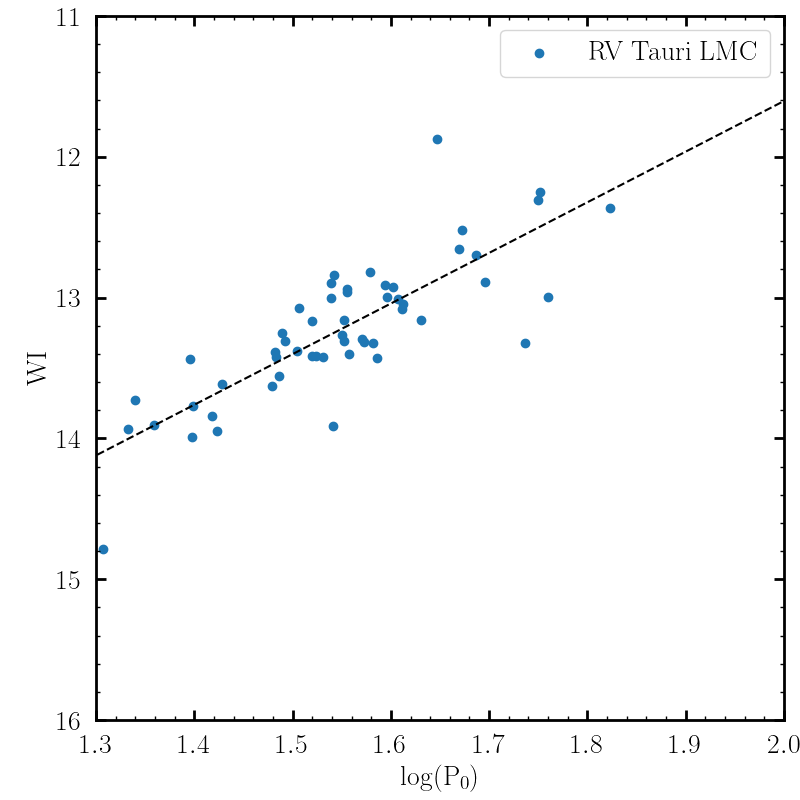}
        \label{subfig:WI_LMC}
    \end{minipage}%
    \begin{minipage}{0.5\textwidth}
        \includegraphics[width=\linewidth]{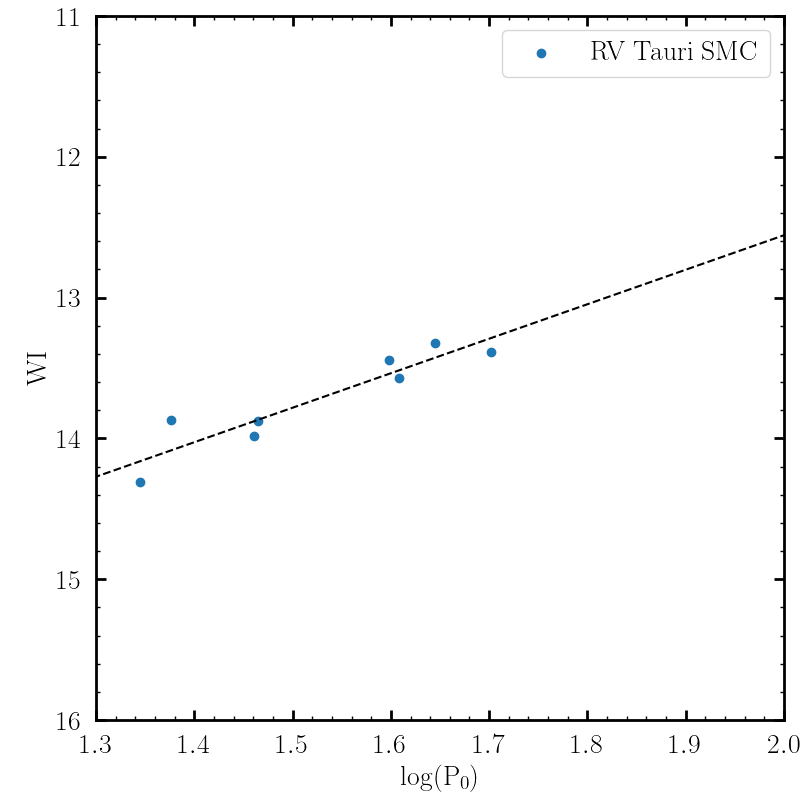}
        \label{subfig:WI_SMC}
    \end{minipage}
    \caption{The reddening-free Wesenheit index plotted with the log ($P_0$) of the RV Tauri stars in the LMC and SMC from \citet{Soszy2015,Soszy2017}.\\ \textit{Note:} We excluded stars from the analysis if their standard deviation exceeded $1\sigma$ of the fit.}
    \label{fig:PLC}
\end{figure*}

We determined the luminosities of the two RV Tauri stars: J005107 and MACHO\,47.2496.8, by adopting the PLC relation calibration method as established by \citet{Manick2018}. The PLC relation was calibrated separately for the LMC and SMC stars using the photometric data in the latest OGLE IV database \citep{Soszy2018}. This relation uses the colour-corrected V-band magnitude known as the Wesenheit index (WI) (\citep{Ngeow2005} and is given by
\begin{equation}
     M_{bol,WI} = m\cdot log(P_0)+c-\mu+BC+2.55(V-I)_0
\end{equation}
where $P_0$ is the observed fundamental pulsation period in days, m is the slope and c is the intercept of the linear regression in the Wesenheit plane (see Figure~\ref{fig:PLC}). We obtained $m = -3.59$ and c = 18.79 for the LMC and $m = -2.45$ and c = 17.45 for SMC respectively. $\mu$ is the distance modulus for the LMC and SMC and have values of 18.49 \citep{Walker2012,Pietrzy2013} and 18.965 \citep{graczyk}, respectively. The value BC is the bolometric correction for each star computed using the relation between BC and effective temperature provided by \citet{Flower1996}. The intrinsic colour $(V-I)_0$ of each star in the SMC and LMC is calculated using the reddening $E(V-I)$ and the observed colour $(V-I)$. The reddening $E(V-I)$ is determined by the conversion of the total reddening $E(B-V)$ derived from the SED (see Section~\ref{sec:luminosity_sed}) using the conversion equation by \citet{Tammann2003} and \citet{Haschke2011}.
\begin{equation}
    E(V-I) = 1.38 E(B-V)
\end{equation}
The derived PLC luminosities (\Lplc) along with their corresponding upper and lower limits ($\Delta$\Lplc) of the RV Tauri targets are displayed in Col. 5 and 6 of Table~\ref{tab:Lumiosity_sample} respectively. To calculate these upper and lower limits for the PLC luminosities ($\Delta$\Lplc), we utilise the corresponding upper and lower limits of the reddening values, as outlined in Col 7 of Table~\ref{tab:Lumiosity_sample}, and also incorporate the error propagation values obtained through the PLC calibration fit.

We once again note that \Lplc\ offers greater precision and reliability when compared to \Lsed, as detailed in the last paragraph of Section~\ref{sec:luminosity}. However, for HD\,158616 we had to consider \Lsed\ as there is no \Lplc\ (this is not an RV Tauri star).

The luminosities of the target stars are presented in Table~\ref{tab:Lumiosity_sample}, and the derived luminosities fall within the typical post-AGB luminosity range. Consequently, we investigate the chemical peculiarities of the target stars, within the context of the post-AGB evolutionary phase. 

\subsection{Initial Mass Estimates}
\label{sec:initial_mass}

The evolution of LIM stars during the post-He-burning life is characterised by the gradual increase in the luminosity, in turn, related to the growth of the mass of the degenerate core, due to the continuous flux of H-free matter, favoured by the CNO nuclear activity. As far as AGB stars are concerned, a classic relation between core mass and luminosity was found by \citet{paczynski}. As stars with different masses evolve through the AGB with different core masses, the luminosity of AGB stars serves as a reliable indicator of the mass of the progenitor. On the chemical side, the surface composition is exposed to changes during the AGB lifetime, primarily connected to the occurrence of several TDU events, which raise the surface carbon, and potentially lead to the formation of carbon stars \citep{iben74}. The increase in the surface carbon takes place in parallel with the growth of the core mass (hence of the luminosity) and is accompanied by the surface \sprocess enrichment. Based on these reasons, the combined knowledge of the luminosity and the surface chemical composition can be tentatively used to identify the progenitors of the sources considered and the evolutionary stage when the AGB evolution was halted, and the contraction to the post-AGB phase began.

\begin{table*}
    \centering
    \caption{Overview of the \Teff, metallicity, \cfe, \sprocess indices, \co\ ratio and initial mass of our target stars.}
        \label{tab:target_ratios}
        \begin{tabular}{ c c c c c c c }
            \hline
            \addlinespace
            Object & \Teff & \feh & \cfe & \sfe & \co & Initial mass range \\
            \addlinespace
            \hline
            \addlinespace
            J005107 & 5768 $\pm$ 85 & -1.57 $\pm$ 0.10 & 1.09 $\pm$ 0.21 & 1.52 $\pm$ 0.20 & 0.58 $\pm$ 0.37 & 0.8-1 \Msun \\
            MACHO\,47.2496.8 & 4900 $\pm$ 250 & -1.50 $\pm$ 0.50 & - & 1.96 $\pm$ 0.20 & $>2$ & 1-1.2 \Msun \\
            HD\,158616 & 7250 $\pm$ 125 & -0.64 $\pm$ 0.12 & 0.47 $\pm$ 0.14 & 0.96 $\pm$ 0.17 & 0.94 $\pm$ 0.22 & 2.5 \Msun \\
            \addlinespace
            \hline
        \end{tabular}
    \begin{center}
    \begin{tablenotes}
     \small
    \item \textbf{Notes:} \Teff, \feh, \cfe, \sfe\ and \co\ of MACHO\,47.2496.8 and HD\,158616 are taken from the high resolution spectroscopic analysis conducted by \citet{Reyniers2007} and \citet{desmedt2016} respectively.\\
    Although an accurate abundance derivation for carbon (\cfe) is absent for MACHO,47.2496.8, it is still categorised as a carbon-enriched object due to its high \co\ and carbon isotopic ratio, as indicated by \citet{Reyniers2007}.
    The \co\ ratio for J005107 should be treated as a lower limit as the precise determination of carbon was not feasible (see Section~\ref{sec:results_J005107}).\\
    The \co\ ratio for MACHO\,47.2496.8 is a lower limit, as the precise determination of carbon and oxygen abundances was not feasible (see \citet{Reyniers2007} for additional details).  
    \end{tablenotes}
    \end{center}  
\end{table*}
\begin{figure}
    \includegraphics[width=\columnwidth]{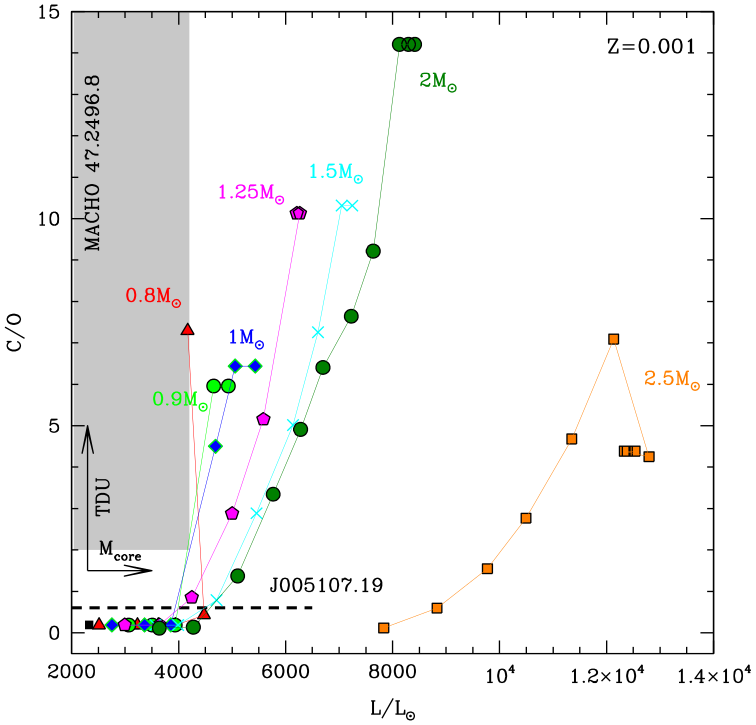}
    \caption{AGB evolution of stars of different mass and metallicity $\rm Z=0.001$, in the $\rm C/O$ vs. luminosity plane. The target stars J005107 and MACHO\,47.2496.8 are marked. See text for more details.}
    \label{fig:ATON1}
\end{figure}
\begin{figure}
    \includegraphics[width=\columnwidth]{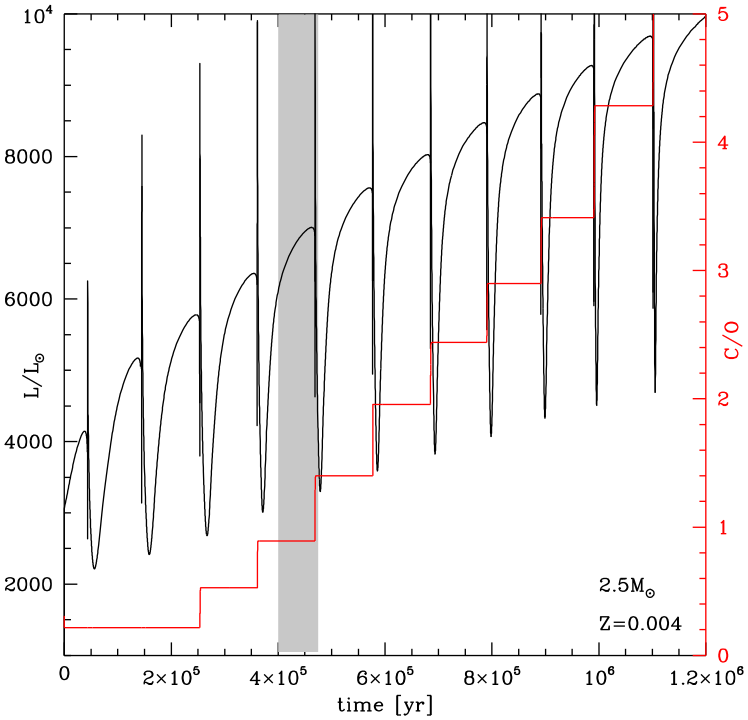}
    \caption{Time variation of the luminosity and the surface $\rm C/O$ of a $\rm 2.5~M_{\odot}$ model star of $Z=0.004$. The grey-shaded region indicates the evolutionary phase of HD\,158616 during which both the luminosity and the \co\ are consistent with those derived from the observations. See text for more details.}
    \label{fig:ATON2}
\end{figure}
In this section, we use ATON code for stellar evolution modelling \citep{ventura98} to estimate the mass of the progenitors of our target stars and their formation epoch. While the ATON evolutionary models are designed for single AGB stars, we explore the potential of employing these single AGB models to characterise the chemical composition of our binary post-AGB target stars, based on the assumption that the photospheric chemical abundance pattern observed in binary post-AGB stars is a reflection of the AGB nucleosynthesis that occurred before the termination of the AGB phase due to the binary interaction. By comparing our observations with predictions from the ATON stellar evolution models, we can indeed confirm whether binary interactions impact the stellar chemical composition. We used evolutionary tracks with an initial metallicity of Z = 0.001 for J005107 and MACHO\,47.2496.8, and Z = 0.004 for HD\,158616. The observationally derived values of \cfe, \sfe, and \co, along with the corresponding initial mass range as determined in this section, are presented in Table~\ref{tab:target_ratios}. Note that the initial mass range indicated here refers to the mass at the beginning of the AGB.

In Figure~\ref{fig:ATON1} we show the AGB evolution of stars of different mass and metallicity $\rm Z=0.001$, in the (surface) $\rm C/O$ vs. luminosity plane. This is similar to the plane used by \citet{devika23} to interpret the sample of Galactic AGBs presented in \citet{Kamath2022}. The single evolutionary sequences run towards the right upper part of the plane, as both luminosity and the surface $\rm C/O$ increase as a consequence of the growth in the core mass and of the effects of TDU. The role of the initial mass is seen in the largest $\rm C/O$ reached, which is seen to be higher the larger the initial mass of the star, as higher mass stars experience more TDU events than their lower mass counterparts, thus reach higher carbon mass fractions in the surface regions \citep{flavia15}. The $\rm 2.5~M_{\odot}$ model star partly escapes from this trend, as the surface carbon enrichment is less efficient in stars with mass close to the threshold to experience hot bottom burning \citep{flavia15}. We also note that the luminosity at which a given value of $\rm C/O$ is reached is sensitive to the initial mass of the star. This is particularly evident for $\rm M > 1.5~M_{\odot}$ (see Figure~\ref{fig:ATON1}), above the threshold mass to start helium burning under quiescent conditions; indeed all the stars undergoing the helium flash develop cores of similar mass. Thus they start the AGB phase with similar luminosities \citep{Ventura2021}.

The dashed horizontal line in Figure~\ref{fig:ATON1} refers to J005107. It is located in correspondence with the lower limit for the surface $\co\sim\!0.58$, while the width indicates the luminosity \Lplc\ along with the upper and lower limits $\Delta$\Lplc\ (see Table~\ref{tab:Lumiosity_sample}). Unfortunately, the latter quantity is very large, of the order of $\rm 5000~L_{\odot}$, thus preventing a tight derivation of the mass of the progenitor. The comparison with the evolutionary tracks shows compatibility with progenitors of $\rm 1-2~M_{\odot}$, corresponding to the formation epoch from 1 Gyr to 3 Gyr ago. This explanation holds provided that the luminosity is above $\rm 4000~L_{\odot}$. Taking into account the recommended value of the base luminosity $\rm 2868~L_{\odot}$ (see Table~\ref{tab:Lumiosity_sample}) poses severe problems in the interpretation of this source, because, as shown in Figure~\ref{fig:ATON1}, no occurrence of TDU events is expected according to the results based on standard stellar evolution modelling in the low luminosity domain. A possible explanation, similar to the one proposed by \citet{devika23} to interpret a few low-luminosity sources in the sample by \citet{Kamath2022}, is that J005107 descends from a low-mass ($\rm 0.8-1~M_{\odot}$) progenitor, which is currently contracting to the blue after losing the external envelope, before the start of the thermal pulses phase. The carbon and \sprocess enrichment would be favoured by the occurrence of deep mixing during the helium flash, similarly to the mechanism invoked by \citet{schwab} to explain the presence of lithium-rich clump stars. A further possibility is that J005107 suffered a late thermal pulse after the beginning of the post-AGB contraction \citep{iben84} so that it is nowadays re-expanding to the red, at a luminosity significantly lower than that characterising the contraction phase. Examples of such an evolution are shown in \citet{blocker95} (see Figure 14) and \citet{tosi22} (see Figure 3).

The grey-shaded region in Figure~\ref{fig:ATON1} refers to MACHO 47.2496.8, based on the lower limit of $\sim 2$ of the surface $\rm C/O$ \citep{Reyniers2007}, and the luminosity \Lplc\ range reported in Table~\ref{tab:Lumiosity_sample}. We find consistency with the predictions from AGB modelling only if the luminosity is close to the upper limit given in Table~\ref{tab:Lumiosity_sample}, as no carbon enrichment is expected at lower luminosities. According to this interpretation MACHO\,47.2496.8 descends from a progenitor of mass in the $\rm 0.8-1~M_{\odot}$ range, which left the AGB after one or two thermal pulses after the C-star stage was reached. We once again note that the values indicated above refer to the mass at the beginning of the AGB. If we consider a typical mass loss during the ascending of the red giant branch of $\rm \sim 0.2~M_{\odot}$, we deduce that the progenitor mass was in the $\rm 1-1.2~M_{\odot}$ range, which corresponds to an age range for MACHO\,47.2496.8 of $3-6$ Gyr.

To study HD\,158616 we use the $Z=0.004$ tracks published in \citet{devika23}. The luminosity range reported in Table~\ref{tab:Lumiosity_sample} rule out very low-mass progenitors, thus shifting the attention to $\rm M > 1.5~M_{\odot}$ stars. In Figure~\ref{fig:ATON2}, we show the time variation of the luminosity and the surface $\rm C/O$ of a $\rm 2.5~M_{\odot}$ model star of $Z=0.004$. The grey-shaded region indicates the evolutionary phase during which both the luminosity and the $\rm C/O=0.94\pm0.22$ \citep{desmedt2016} are consistent with those derived from the observations. Based on these results, we propose that HD\,158616 descends from a $\rm 2.5~M_{\odot}$ progenitor, formed around half Gyr ago, which left the AGB after experiencing a few thermal pulses, two out of which after the first TDU event. The estimated mass interval for the progenitor is pretty narrow in this case since the initial mass vs luminosity trend is very tight in the $\rm M > 2~M_{\odot}$ domain (see Figure~\ref{fig:ATON1}).

\section{Discussion}
\label{sec:discussion}
In the following subsections, we employ the obtained stellar parameters, chemical abundances, derived luminosities, and initial masses to explore the chemical peculiarity of our target stars (Section~\ref{sec:chemical_peculiarity}). We also investigate whether the observed \sprocess enrichment of our targets has an extrinsic or intrinsic nature (see Section~\ref{sec:intrinsic_extrinsic}).

\subsection{Investigating the chemical peculiarity of Targets}
\label{sec:chemical_peculiarity}
\begin{figure*}
    \begin{subfigure}{0.33\textwidth}
        \centering
        \includegraphics[width=\linewidth]{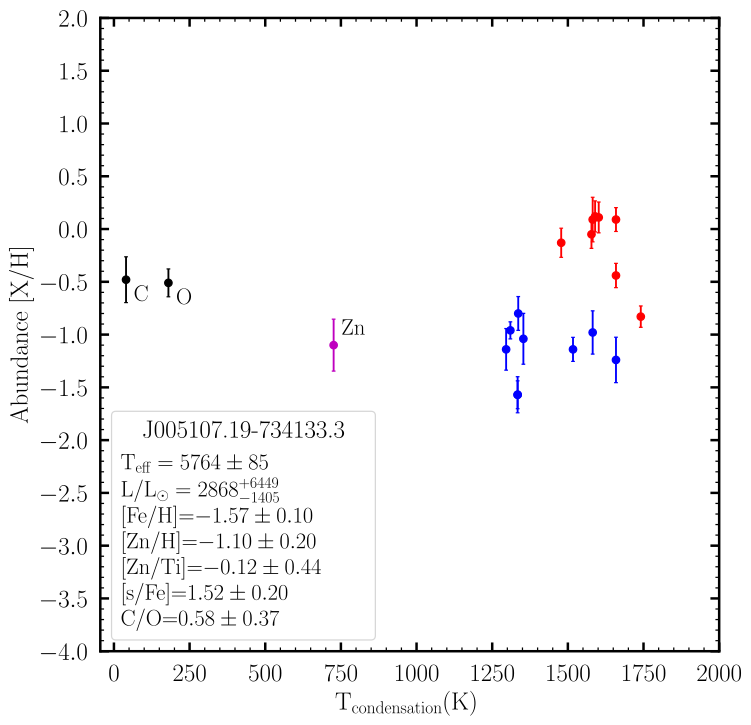}
    \end{subfigure}%
    \begin{subfigure}{0.33\textwidth}
        \centering
        \includegraphics[width=\linewidth]{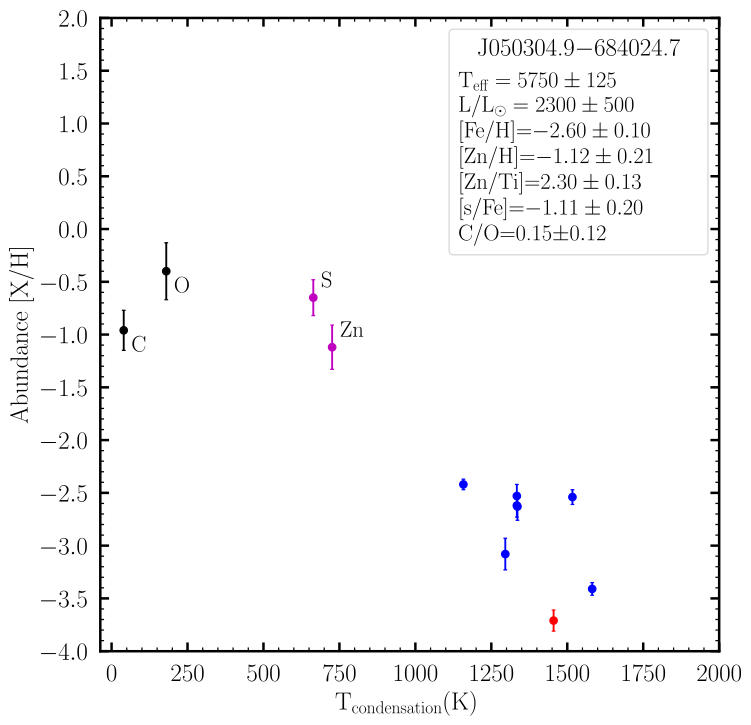}
    \end{subfigure}%
    \begin{subfigure}{0.33\textwidth}
        \centering
        \includegraphics[width=\linewidth]{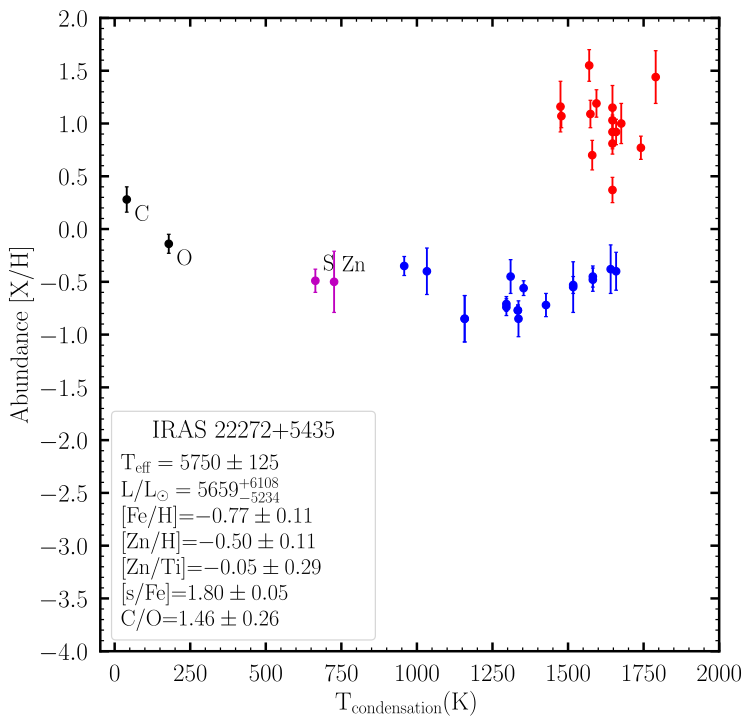}
    \end{subfigure}%
    
    \begin{subfigure}{0.33\textwidth}
        \centering
        \includegraphics[width=\linewidth]{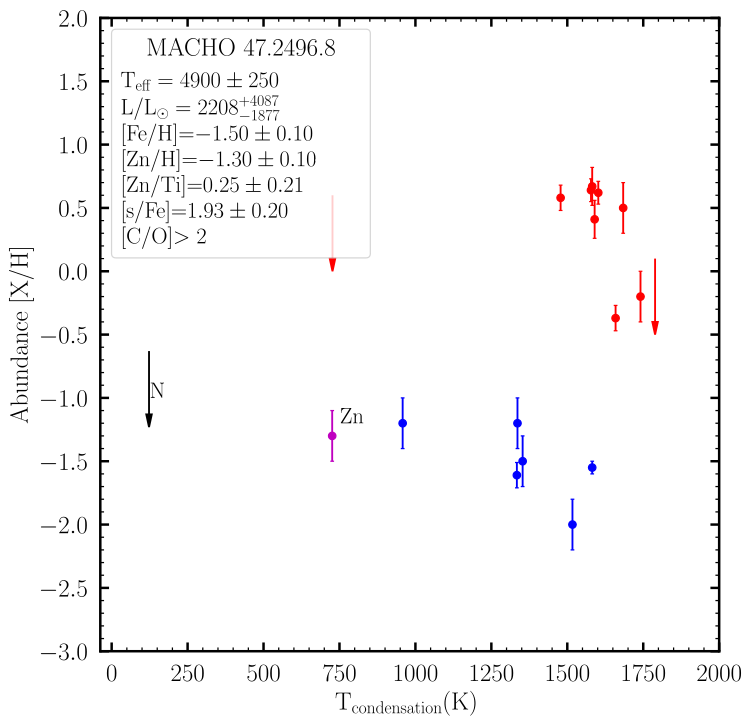}
    \end{subfigure}%
    \begin{subfigure}{0.33\textwidth}
        \centering
        \includegraphics[width=\linewidth]{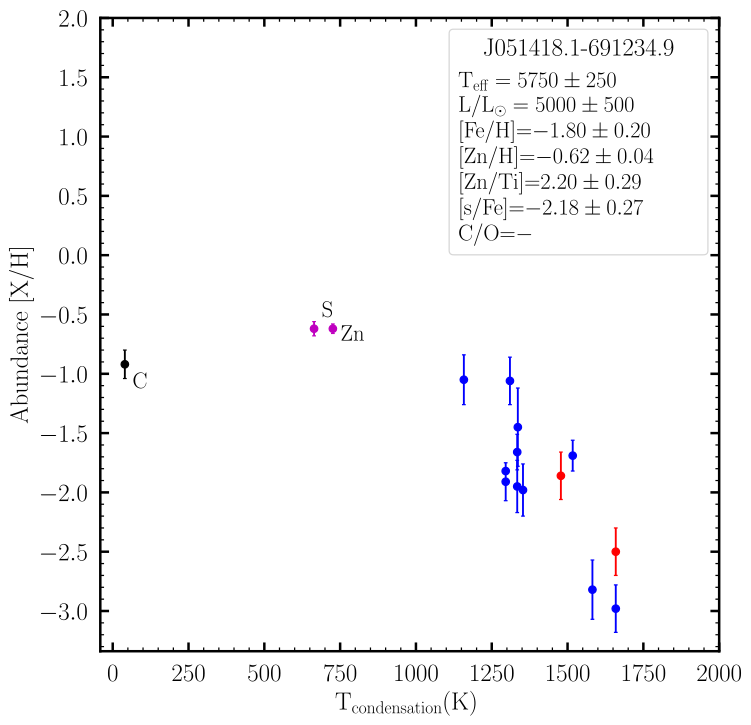}
    \end{subfigure}%
    \begin{subfigure}{0.33\textwidth}
        \centering
        \includegraphics[width=\linewidth]{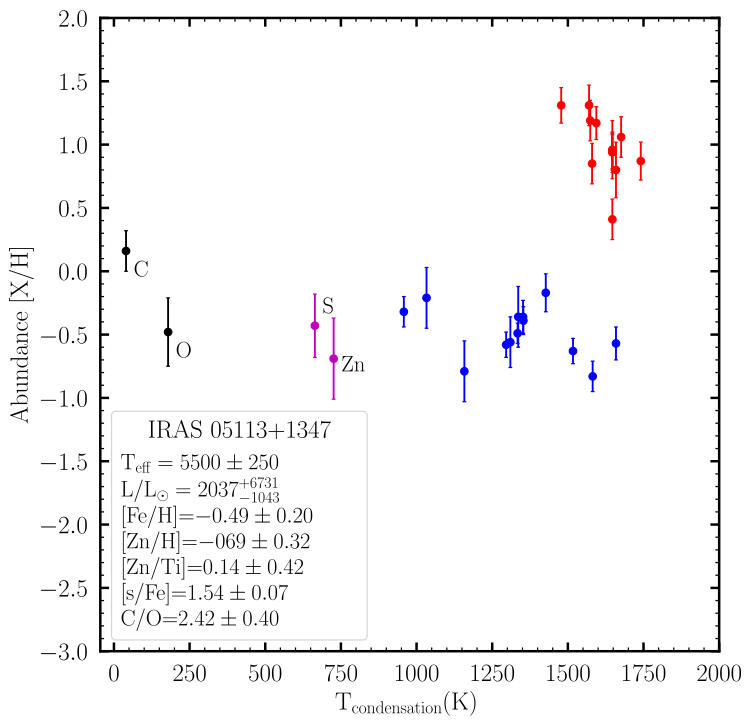}
    \end{subfigure}%
    
    \begin{subfigure}{0.33\textwidth}
        \centering
        \includegraphics[width=\linewidth]{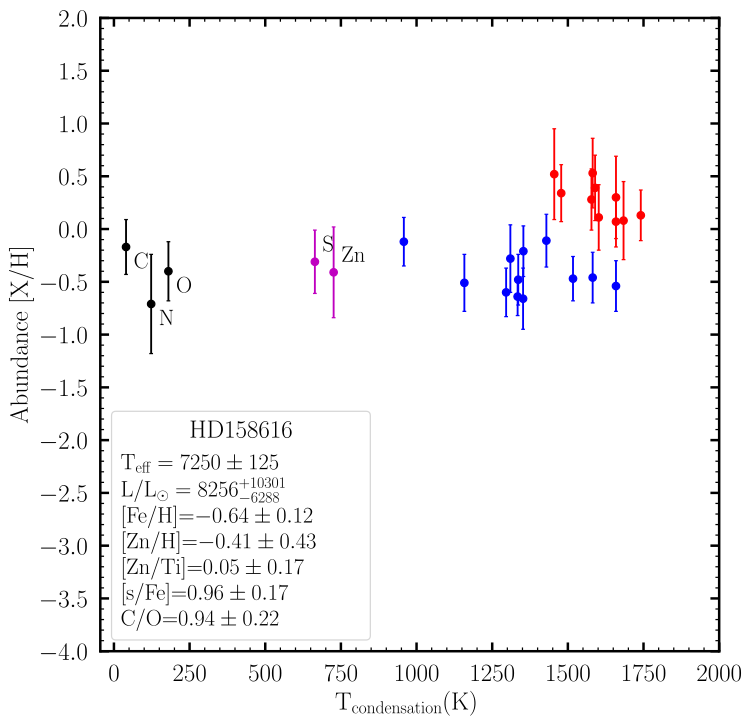}
    \end{subfigure}%
    \begin{subfigure}{0.33\textwidth}
        \centering
        \includegraphics[width=\linewidth]{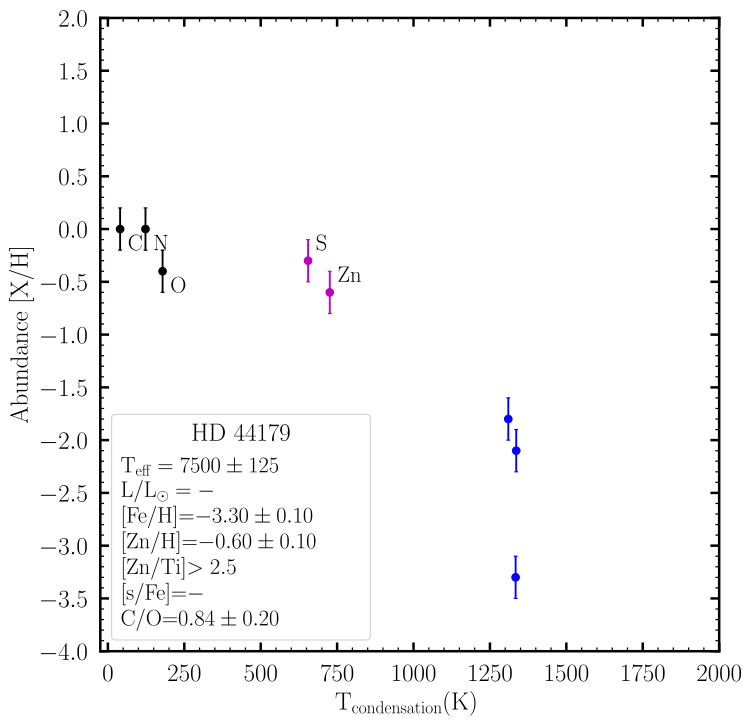}
    \end{subfigure}%
    \begin{subfigure}{0.33\textwidth}
        \centering
        \includegraphics[width=\linewidth]{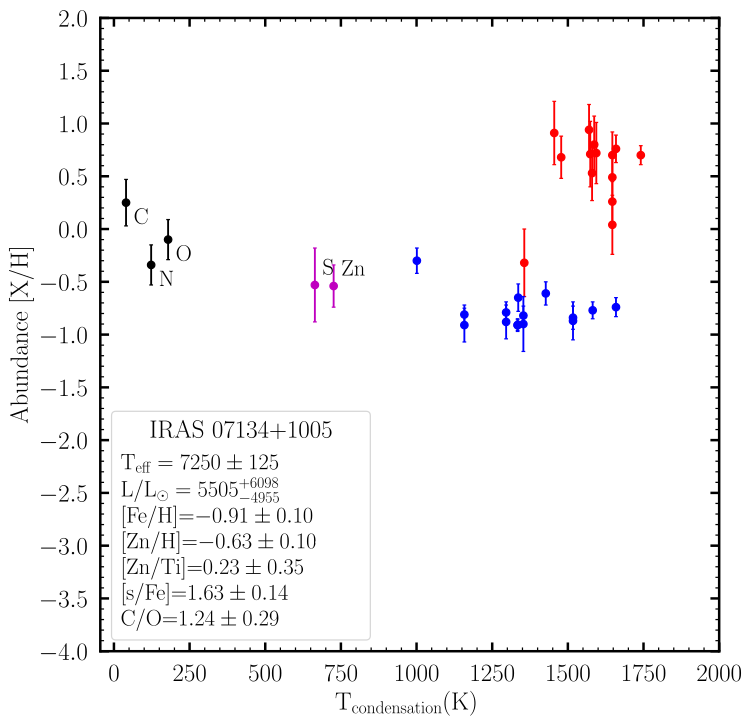}
    \end{subfigure}%
   
    \caption{Comparison plots illustrating the abundance pattern in terms of element over hydrogen (\xh) ratios as a function of condensation temperature (\Tcondensation) \citep[taken from][]{Lodders2003} of the target stars with post-AGB binary stars and post-AGB single stars from the literature that have similar stellar parameters. The stellar parameters along with other necessary parameters are provided in the legend. To distinguish between the different elements, we continue to use the same colour coding and symbols as before. The black colour data points represent CNO elements, the blue represents Fe peak elements, the magenta represents Zn and S, and the red represents \sprocess elements.\\
    \textit{Left:} \xh\ Vs \Tcondensation\ of target stars.\\
    \textit{Middle:} \xh\ Vs \Tcondensation\ of post-AGB binary stars from \citet{kamath19}, \citet{Gielen2009} and \citet{Waelkens1992} with similar stellar parameters of the target sample in the respective left panel.
    \textit{Right:} \xh\ Vs \Tcondensation\ of post-AGB single stars from \citet{desmedt2016} that has similar stellar parameters of the target sample in the respective left panel. 
    }
    \label{fig:comparisonPlotsCondensation}
\end{figure*}

We compared the chemical compositions of our target stars to appropriate binary post-AGB stars and single post-AGB stars with similar stellar parameters: \Teff, \logg\ and \feh\ (see Figure~\ref{fig:comparisonPlotsCondensation} legends). This comparative analysis is aimed at discerning whether there exists a composite signature reflecting both chemical depletion, commonly observed in binary post-AGB stars, and \sprocess enrichment, typically seen in low-mass single post-AGB stars. Identifying both chemical patterns concurrently might suggest the likelihood of \sprocess enrichment, consequently leading to the observed chemical depletion pattern. 
 
To investigate any hidden signatures of chemical depletion in our targets, we make use of the ratio \znti\ (see \citet{kamath19} for more details). Since Ti is likely to be more strongly depleted than Zn, \znti\ is an ideal indicator of chemical depletion in O-rich environments. We note that the abundance of CNO elements may undergo changes as a result of prior AGB nucleosynthesis and dredge-up. Consequently, understanding the true effect of depletion on these elements poses a challenge and falls outside the scope of this study. Our analysis mainly focuses on examining the effect of depletion on the refractory elements.

Figure~\ref{fig:comparisonPlotsCondensation} shows the abundance of the element over hydrogen (\xh) as a function of the equilibrium condensation temperature (\Tcondensation) for a solar photospheric composition, which is oxygen-rich, \citep[taken from][]{Lodders2003} for the individual target stars (left panels) and their comparison binary stars (middle panels) and their comparison single stars (right panels). The stellar parameters (\Teff, \logg\ and \feh), luminosities, \znti, \sfe\ and \co\ for each of the stars are listed in their respective legends. 

For J005107 we use J050304.9-684024.7 \citep{kamath19} as our comparison binary post-AGB star and IRAS\,22272+5435 \citep{desmedt2016} as our comparison single post-AGB star. For MACHO\,47.2496.8 we use J051418.1-691234.9 \citep{Gielen2009} as our comparison binary post-AGB star and IRAS\,05113+1347 \citep{desmedt2016} as our comparison single post-AGB star. For HD\,158616 we use HD\,44179 \citep{Waelkens1992} as our comparison binary post-AGB star and IRAS 07134+1005 \citep{desmedt2016} as our comparison single post-AGB star.

From Figure~\ref{fig:comparisonPlotsCondensation}, it becomes evident that the chemical patterns of our targets align consistently with their comparison single post-AGB stars (as expected for post-TDU objects), indicating that our targets show notable enrichment in \sprocess elements with no signs of chemical depletion unlike their comparison binary post-AGB stars. The value of $\znti\,<\!0.5$ for our target stars (see legends of Figure~\ref{fig:comparisonPlotsCondensation}) further confirms the absence of a photospheric chemical depletion. Therefore, we dismiss the possibility of an \sprocess enrichment following a chemical depletion. 

As mentioned in Section~\ref{sec:intro}, the chemical depletion pattern observed in binary post-AGB stars with circumbinary discs is attributed primarily to the disc-binary system interaction. Since our target stars do not show the expected signatures of chemical depletion, it appears that the disc-binary interaction inducing chemical depletion failed in our systems. To explain this anomaly, it is worth noting that the majority of the binary post-AGB stars that display photospheric chemical depletion have a photospheric chemical composition that reflects a $\co\,<\!1$ dex (the majority having a $\co\,\lessapprox\!0.6$) and circumbinary disc chemistry that is O-rich \citep[see][and references therein]{Gielen2011,Cava2023,Mass2005}. Accordingly, the condensation temperatures typically utilised (as also in this study, as mentioned above) are based on a solar photospheric composition characterised by an O-rich nature \citep{Lodders2003}. However, our target stars display a photospheric chemical composition that is enhanced in carbon with $\co\,>\!1$ dex, \cfe\ ranging between 0.5\,-\,1.0 dex (dependant on metallicity) and \sprocess elements with $\sfe\,>\!1$ dex. This points to a possibility of the circumbinary disc around our target stars possessing a C-rich dust chemistry rather than the atypical O-rich dust chemistry. Unfortunately, we do not have observations of the circumstellar disc chemistry of our target stars to verify their C-rich nature. However, it is well known that the mechanisms governing processes such as molecular formation and dust condensation are likely to exhibit significant differences \citep{Dell2021} resulting in likely variations to chemical depletion efficiencies in C-rich environments when compared to O-rich environments. Consequently, dust condensation temperatures based on an O-rich disc composition, as used in this study, will not be suitable for our targets. To fully grasp the impact of C-rich circumstellar chemistry on chemical depletion in binary post-AGB stars, we would also need condensation temperatures specific to a C-rich environment. Unfortunately, these estimates are not currently available, making the qualitative investigation of this aspect outside the scope of our study.

\begin{figure}
    \includegraphics[width=\columnwidth]{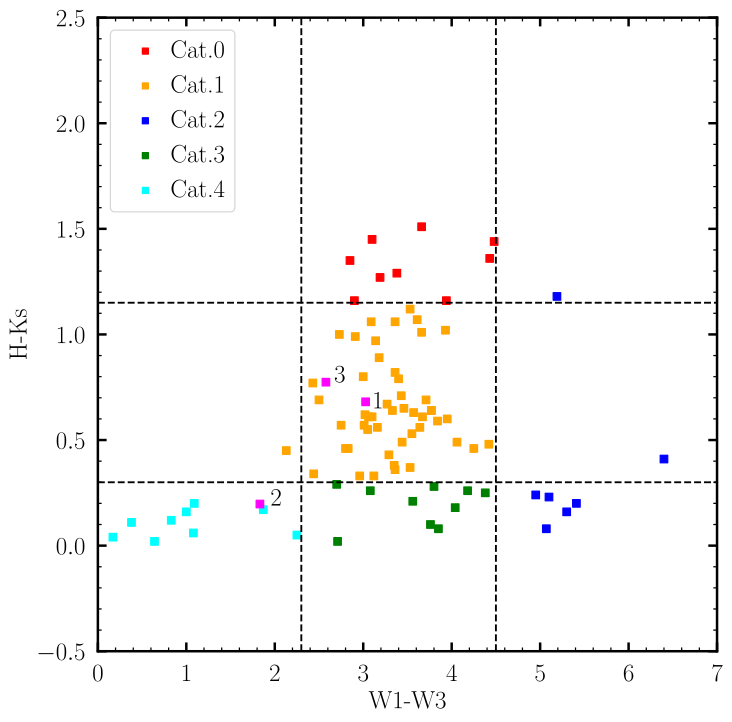}
    \caption{Wise colour-colour diagram for the target stars along with the stars studied in \citet{Kluska2022}. The dashed black lines represent the boundaries between the different categories defined in \citet{Kluska2022}. The targets of this study are given in magenta colour and are numbered according to their position in Table~\ref{tab:sample} for reference.}
    \label{fig:CategoryPlot}
\end{figure}

To examine potential variations in the disc morphologies of the circumbinary discs around our target stars, we utilise the colour-colour diagram as introduced by \citet{Kluska2022}, which is based on the characteristics of the IR excess. This approach aids in identifying the circumbinary disc type associated with our targets. Figure~\ref{fig:CategoryPlot} displays the IR colour-colour diagram of the target stars along with the stars studied in \citet{Kluska2022}. J005107 and HD\,1586161 exhibit a "full-disc" model. This implies continuous dust extension from the dust sublimation radius to the outer edge. In contrast, MACHO\,47.2496.8 features a "transition-disc" model, indicating a lack of dust within a radius significantly larger than the theoretical dust sublimation radius (for further insights, refer to \citet{Kluska2022}). These observations clarify that the circumbinary disc of our target stars closely resembles that of depleted binary post-AGB stars, presenting either as a full-disc or a transition-disc. Therefore, there is no apparent deviation in the circumbinary disc morphology or characteristics from the typical traits observed in a binary post-AGB star. This indicates that the observed chemical peculiarity of the target stars is not attributed to their disc structure.

\subsection{Extrinsic or Intrinsic Enrichment?}
\label{sec:intrinsic_extrinsic}
The observed enrichment in \sprocess (and carbon) elements within our target stars is likely to be due to three possibilities: 1) extrinsically from the binary companion, 2) intrinsically through nucleosynthesis processes occurring within the star during its AGB phase, and 3) from the host galaxy, through intrinsic enrichment from the chemical composition of the interstellar medium (ISM) from which the star originated. We investigate each of these possibilities in detail.

\textbf{\textit{Extrinsic enrichment:}} 
Observational investigations of low-mass binary stars, such as Barium stars and Carbon-Enhanced Metal-Poor (CEMP) stars, indicate that the observed carbon and \sprocess enrichment in their photospheres results from interactions with their evolved companions (AGB or white dwarf (WD)) \citep[e.g.,][]{Jorissen2019,Escorza2023,Goswami2023}. The companion star, having undergone or undergoing AGB nucleosynthesis, transfers carbon and \sprocess elements to the photosphere of these stars, making them extrinsically enriched. We investigated a similar possibility for our targets. However, previous studies of Galactic binary post-AGB stars \citep[see][e.g.]{oomen18} show that most of these stars do not have a dim WD as a companion. We present some of the arguments favouring this hypothesis below.

The first is a statistical argument. It is supported by the robust analysis of the companion mass ($\rm M_2$) distribution conducted by \citet{oomen18}. From their analysis, it was found that the companion mass distribution function (see Figure~4 of \citet{oomen18}) peaks at $\sim\!1\Msun$. To explore the possibility of a WD companion population, \citet{oomen18} fitted a double Gaussian profile to the mass distribution. However, this adjustment did not result in a better fit, and the mass distribution shifted towards higher masses rather than peaking at lower masses comparable to WDs. Consequently, they refrained from incorporating the WD population into the mass distribution. Additionally, data from the Gaia DR2 catalogue indicates that the observed distribution of WD mass peaks at $0.6\Msun$ \citep{Jim2018}. Therefore, the cumulative mass-function distribution fit by \citet{oomen18} does not convincingly support the possibility that the majority of the binary post-AGB stars have a WD companion as the mass distribution of companion masses does not have a peak around $0.6\Msun$.

The second is an observational argument. The majority of the binary post-AGB systems have jets launched from their circum-companion \citep[e.g.][]{Bollen2017,Bollen2021,Bollen2022}. The velocities of these jets can be used to constrain the nature of the jet-launching object, i.e. the companion in these binary systems. The velocity of these jets ranges from $\sim\!150$ to 400 \kms\ \citep{Bollen2021,Bollen2022}. The escape velocity of a WD ($\sim\!5000$ \kms) is about one order of magnitude larger than the jet velocities determined for the post-AGB binary systems. On the other hand, the jet velocities from the binary post-AGB system are all around the escape velocity of MS stars ($\sim\!100$ to 1000 \kms). This suggests that the companion of the majority of the binary post-AGB stars are MS. 

The third argument is based on the observations of binary post-AGB stars by \citet{oomen18}, where they did not detect any symbiotic activities in the spectra. This lack of symbiotic activities, when considered alongside the two aforementioned points, serves as an additional indication supporting the absence of a compact object, such as a WD, within the binary system. 

All the above-mentioned arguments are, therefore, a strong indication that the companions in the majority of the binary post-AGB systems are likely to be MS stars rather than WDs. With regard to the targets in our study, the absence of orbital parameter data for two targets (J005107 and MACHO\,47.2496.8) limits the extent of our argumentation. However, the confirmed binary star HD 158616 in our study has a detected mass function of 0.022\Msun, which translates to a companion mass of 0.26\Msun\ assuming a mass of 0.6\Msun\ for the primary star and an inclination of 75 degrees (too small companion mass to be a WD at 75 degrees inclination) \citep[see][]{oomen18}. As the light curve is not affected by variable reddening, the inclination may be larger, leading to an even higher companion mass. Additionally, HD\,158616 has an orbital period of $363.3$ days, which means that the full orbital phase coverage is difficult to obtain given the observations are always covering the same phase. Yet we have accumulated enough data to show that the jets as observed in this object ($\sim\!150$ \kms) do not trace the escape velocity of a WD but that of a MS star (see Figure~\ref{fig:hd158616_dynamic}). Therefore, based on HD158616, and extrapolating the arguments observed for the majority of the binary post-AGB stars to our sample objects where orbital data is unavailable, we can conclude that extrinsic enrichment is highly unlikely for our targets. 

\textbf{\textit{Intrinsic enrichment:}} To investigate the possibility of intrinsic enrichment, we used ATON code for stellar evolution modelling \citep{ventura98} to trace the nucleosynthetic history of our target stars (see Section~\ref{sec:initial_mass} for more details). We once again note that while the ATON evolutionary models are designed for single AGB stars, we explore the potential of employing these models to predict $\rm C/O$ for each of the target stars (as tracers of TDU). We then compared the theoretically predicted \co\ with the observed $\rm C/O$ of our target stars. The results from the models predict that our target stars have experienced thermal pulses during their AGB phase (see Figure~\ref{fig:ATON1} and Figure~\ref{fig:ATON2}, respectively). For J005107, this result is explained using the upper limit luminosity range \Lplc\ (see Table~\ref{tab:Lumiosity_sample}). We note that the observed O abundance of J005107 ($\ofe\ = 1.06\pm0.21$ dex) is also mainly intrinsic and influenced by the dredge-up in the prior AGB phase. During the TDU, even if the C enrichment is the most outstanding consequence, some O enrichment occurs too. Especially in the metal-poor domain, the increase in \ofe\ is relevant due to the low oxygen content of the gas from which the star was formed. Concerning the Z = 0.001 models used in this paper for J005107 (see Section~\ref{sec:initial_mass}), we find that the largest increase in \ofe\ is achieved during the evolution of the 2.5\Msun\ model star, for which \ofe\ increases by 0.9 dex. For lower initial masses we still find O enrichment, down to 0.2 dex for the stars of initial mass around 0.9\Msun.

Consequently, the above arguments show the likely occurrence of AGB nucleosynthesis and hence strongly suggest the likelihood of an intrinsic carbon and \sprocess enrichment of our targets (see Section~\ref{sec:initial_mass} for more details).

\textbf{\textit{Inherited enrichment from the host galaxy:}} We also investigated whether the observed \sprocess enrichment of the target stars originated from the initial chemical composition of the ISM from which it was formed. To assess this, we compared the local initial \sprocess abundances of the SMC, LMC and the Galaxy. The initial \sprocess abundances of the SMC were determined by comparing the \sprocess abundances of SMC red giant stars \citep{Mucciarelli2023}. For the LMC, initial \sprocess abundances were derived from the \sprocess abundances of LMC field red giant stars \citep[][and references therein]{VanderSwaelmen2013}. In the case of the Galaxy, the initial \sprocess abundances were determined by comparing the \sprocess abundances of F and G dwarfs within the Galactic discs, as reported by \citet{Simmerer2004}, \citet{Bensby2005}, \citet{Brewer2006} and \citet{Reddy2003,Reddy2006}. Based on the above comparisons, it becomes evident that the standard enrichment of \sprocess elements in the SMC, LMC, and the Galaxy is typically $\sim\,\sfe\,\leq\!0.5$ dex. V453\,Oph, a post-AGB star in the Galaxy, exhibits a mild \sprocess enrichment which is attributed to the initial inherited \sprocess enrichment from the Galaxy \citep{Deero2005}. Drawing insights from the abundance analysis of V453\, Oph \citep{Deero2005}, we conclude that the mild \sprocess enrichment of the order of $\sfe\ = 0.50\pm0.21$ dex can reflect the inherited \sprocess abundance of the Galaxy at its birthplace, rather than being a result of its internal nucleosynthesis (see \ref{sec:appendix:v453oph} for additional details regarding V453\,Oph). Additionally, it is worth noting that, unlike our targets, V453\,Oph exhibits a very metal-poor composition, with $\feh\ = -2.23\pm0.12$ dex, where the inherited galactic enrichments are typically more prominent. Consequently, for our post-AGB target stars, the impact of inherited galactic enrichment (i.e., a maximum of $\sim\,\sfe\,\leq\!0.5$ dex) cannot solely explain the strong \sprocess enrichment observed (i.e., $\sfe\,\gtrapprox\!1$ dex).

Therefore, the above arguments point to the possibility that the observed carbon and \sprocess enrichment of our target stars is most likely intrinsic in nature. This implies that the observed carbon and \sprocess enrichment is likely to be a result of the AGB nucleosynthesis that occurred before the termination of the AGB phase via binary interaction. However, what remains unclear is that unlike the majority of binary post-AGB which display a disc-binary driven photospheric chemical depletion, our target star's chemical depletion seems to have failed, causing retention of the signature of the nucleosynthesis that occurred during and before the AGB evolution of these objects. 

\section{Summary and Conclusions}
\label{sec:conclusion}
In this study, we present detailed stellar parameter and chemical abundance studies of a subset of chemically peculiar binary post-AGB stars (disc-sources) in the Galaxy and the MCs that exhibit a carbon and \sprocess enrichment in their photospheric chemical composition, contrary to the commonly observed photospheric chemical depletion typically observed in binary post-AGB stars. 

Our investigation into the carbon and \sprocess enrichment observed in our target stars has provided valuable insights. By considering extrinsic, intrinsic, and inherited galactic enrichment scenarios, we have effectively ruled out extrinsic and inherited galactic \sprocess enrichment as explanations for the observed overabundance of carbon and \sprocess elements in our target stars.

Based on our observationally derived stellar parameters and chemical abundances combined with predictions from dedicated ATON stellar evolutionary models (Ventura et al., 1998), we propose that the carbon and \sprocess enrichment in our target stars is predominantly intrinsic in nature. Our target stars display an abundance pattern consistent with expectations for a post-TDU system (as depicted in Figure~\ref{fig:comparisonPlotsCondensation}) with $\co\,>\!1$ dex, \cfe\ ranging between 0.5\,-\,1.0 dex and $\sfe\,>\!1$ dex. This photospheric chemical composition suggests the presence of a circumstellar disc with C-rich chemistry. The differences in molecular formation and dust condensation processes that occur in C-rich versus O-rich circumstellar environments (atypical of the majority of binary post-AGB stars displaying photospheric chemical depletion) strongly indicate potential variations in chemical depletion efficiencies. Furthermore, the standard dust condensation temperatures, based on O-rich compositions, are not suitable for our analysis. However, the absence of observations regarding the circumstellar dust chemistry of our targets, coupled with the lack of condensation temperature estimates specific to C-rich environments, limits our ability to confirm this hypothesis.

Conducting observational studies on the circumbinary disc chemistry of our target stars, especially in comparison to other binary post-AGB stars displaying chemical depletion, is crucial. This effort will uncover potential peculiarities in our target stars and address gaps in our current understanding of the disc-binary interaction that leads to varying efficiencies and/or the lack of chemical depletion in binary post-AGB systems.

\begin{acknowledgement}
MM acknowledges the financial support provided by the International Macquarie Research Excellence Scholarship (iMQRES) program for the duration of this research. MM, DK, and MM also acknowledge the ARC Centre of Excellence for All Sky Astrophysics in 3 Dimensions (ASTRO 3D). P.V. acknowledges the support received from the PRIN INAF 2019 grant ObFu 1.05.01.85.14 (“Building up the halo: chemodynamical tagging in the age of large surveys”, PI. S. Lucatello). HVW acknowledges support from the Research Council, KU Leuven under grant number C14/17/082. This study is based on observations collected with the Very Large Telescope at the ESO Paranal Observatory (Chile) of program numbers 099.D-0536, 074.D-0619 and 094.D-0067.
\end{acknowledgement}



\paragraph{Data Availability Statement}
The data underlying this article is made available online.

\printendnotes
\bibliography{mnemonic,meghna}

\appendix
\section{Spectroscopic Analysis of MACHO 47.2496.8 and HD 158616}
\label{sec:appendix:analysis2targets}
We performed a similar atmospheric parameter analysis and abundance analysis as described in Section~\ref{sec:atmos_param} and Section~\ref{sec:abund_analysis} respectively, for MACHO\,47.2496.8 and HD\,158616.

The results of the atmospheric parameters are presented in Table~\ref{tab:sample} under the title "This study". The results of the abundance analysis are presented in Table~\ref{tab:appendix:abund_MAcho47&HD158616}. We note that since the derived atmospheric parameters and abundances of MACHO\,47.2496.8 and HD\,158616 align closely with values reported in the literature (see Table~\ref{tab:sample} for atmospheric parameter details and Table~\ref{tab:appendix:abund_MAcho47&HD158616} for abundance details), we opted to adopt the literature values of both atmospheric parameters and abundances for the rest of our analysis.

\begin{table*}
    \centering
    \caption{Spectroscopically determined abundance results for MACHO\,47.2496.8 and HD\,158616 along with their literature values.}
        \label{tab:appendix:abund_MAcho47&HD158616}
        \begin{tabular}{ c c c c c c }
            \hline
            \addlinespace
            \multicolumn{2}{ c }{} & \multicolumn{2}{c }{MACHO\,47.2496.8} & \multicolumn{2}{c }{HD\,158616}\\ 
            \addlinespace
            \hline
            Ion & Z & \xfe & $\xfe_{Literature}$ & \xfe & $\xfe_{Literature}$ \\
            \addlinespace
            \hline
            \addlinespace
            C I & 6 & - & - & 0.51$\pm$0.20 & 0.47$\pm$0.14 \\
            N I & 7 & - & <0.88$\pm$0.20 & - & -0.07$\pm$0.35 \\
            O I & 8 & - & - & 0.19$\pm$0.14 & 0.24$\pm$0.16 \\
            Na I & 11 & 0.28$\pm$0.21 & 0.30$\pm$0.20 & 0.47$\pm$0.22 & 0.52$\pm$0.11 \\
            Mg I & 12 & 0.32$\pm$0.20 & 0.30$\pm$0.20 & 0.19$\pm$0.11 & 0.16$\pm$0.12 \\
            Si I & 14 & - & - & 0.37$\pm$0.26 & 0.36$\pm$0.20 \\
            S I & 16 & - & - & 0.33$\pm$0.25 & 0.33$\pm$0.18 \\
            Ca I & 20 & -0.30$\pm$0.20 & -0.50$\pm$0.20 & 0.16$\pm$0.23 & 0.17$\pm$0.09 \\
            Sc II & 21 & - & - & 0.13$\pm$0.17 & 0.10$\pm$0.12 \\
            Ti I & 22 & -0.09$\pm$0.20 & -0.05$\pm$0.05 & 0.20$\pm$0.19 & 0.18$\pm$0.12 \\
            V II & 23 & - & - & 0.54$\pm$0.15 & 0.53$\pm$0.13 \\
            Cr I & 24 & - & - & 0.05$\pm$0.19 & 0.04$\pm$0.11 \\
            Mn II & 25 & - & - & 0.18$\pm$0.21 & 0.13$\pm$0.15 \\
            Fe I & 26 & -0.05$\pm$0.14 & -0.11$\pm$0.10 & 0$\pm$0.18 & 0$\pm$0.06 \\
            Co I & 27 & - & - & -0.23$\pm$0.20 & -0.20$\pm$0.17 \\
            Ni I & 28 & 0.05$\pm$0.01 & 0$\pm$0.20 & 0.36$\pm$0.19 & 0.43$\pm$0.12 \\
            Zn I & 30 & 0.17$\pm$0.20 & 0.20$\pm$0.20 & 0.29$\pm$0.21 & 0.23$\pm$0.31 \\
            Y II & 39 & 1.17$\pm$0.24 & 1.13$\pm$0.10 & 0.72$\pm$0.12 & 0.71$\pm$0.12 \\
            Zr II & 40 & 1.37$\pm$0.20 & 1.30$\pm$0.20 & 0.78$\pm$0.25 & 0.77$\pm$0.12 \\
            Ba II & 56 & - & - & 1.20$\pm$0.22 & 1.16$\pm$0.31 \\
            La II & 57 & 2.10$\pm$0.24 & 2.14$\pm$0.09 & 0.86$\pm$0.11 & 0.92$\pm$0.17 \\
            Ce II & 58 & 2.12$\pm$0.21 & 2.08$\pm$0.10 & 1.08$\pm$0.11 & 0.98$\pm$0.15 \\
            Pr II & 59 & 2.19$\pm$0.20 & 2.17$\pm$0.15 & 1.07$\pm$0.19 & 1.17$\pm$0.21 \\
            Nd II & 60 & 2.16$\pm$0.11 & 2.12$\pm$0.09 & 0.75$\pm$0.23 & 0.75$\pm$0.19 \\
            Sm II & 62 & 2.08$\pm$0.13 & 1.91$\pm$0.15 & 0.98$\pm$0.17 & 1.03$\pm$0.19 \\
            Lu II & 71 & - & - & 0.89$\pm$0.18 & 0.94$\pm$0.27 \\
            Hf II & 72 & 2.12$\pm$0.20 & 2.00$\pm$0.20 & 0.76$\pm$0.20 & 0.72$\pm$0.25 \\
            W I & 74 & \textit{<1.71$\pm$0.20} & <1.60$\pm$0.20 & - & - \\
            Pb I & 82 & \textit{<2.17$\pm$0.20} & <2.10$\pm$0.20 & - & - \\
            \addlinespace
            \hline
        \end{tabular}
    \begin{center}
    \begin{tablenotes}
     \small
    \item \textbf{Notes:} The detected ions and their corresponding atomic number (Z) are listed in columns 1 and 2, respectively. \xfe\ is the element-over-iron ratio along with the total uncertainty (see Section~\ref{sec:abund_analysis} for more details on error estimations). The abundances in italics are derived using the SSF method. The literature values are taken from \citet{desmedt2016} and \citet{Reyniers2007} respectively. 
    \end{tablenotes}
    \end{center}  
\end{table*}

\section{Target details of V453 Oph}
\label{sec:appendix:v453oph}

\begin{figure}
      \includegraphics[width=\columnwidth]{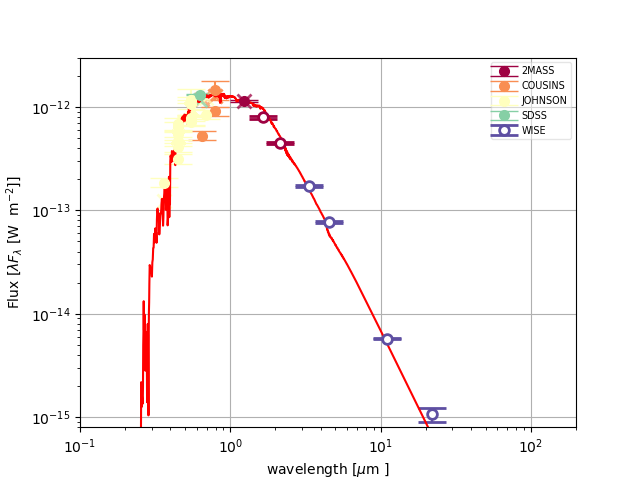}
    \caption{SED of V453\,Oph. The data points represent the dereddened photometry. The red line represents the best-fitting scaled model atmosphere (see Section~\ref{sec:sed} for details).}
    \label{fig:SED_v453oph}
\end{figure}
\begin{figure}
      \includegraphics[width=\columnwidth]{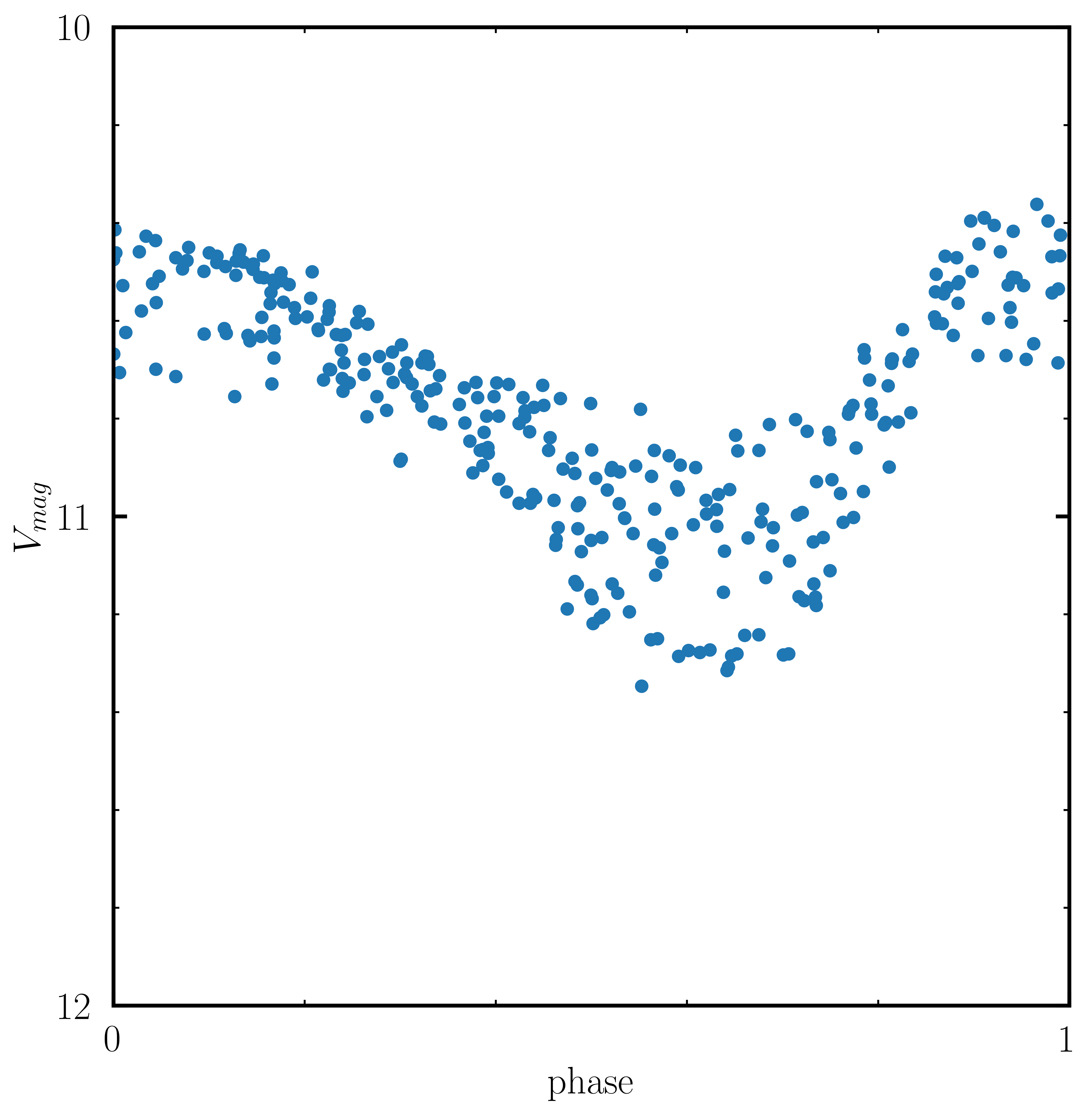}
     \caption{The phased LC of the V453\,Oph. The fundamental pulsation period used to phase the LC is 40.52 days. The LCs are scattered due to their semi-regular nature; they usually show considerable variations from cycle to cycle. This behaviour is typically more pronounced for the longer-period RV Tauri stars. The photometric data for V453\,Oph is taken from ASAS-SN Variable Stars Database II \citep{Shappee2014,Jayasinghe2019}.}
   \label{fig:light_Curve:v453oph}
\end{figure}

V453\,Oph has been classified as a post-AGB star based on its high luminosity \citep{Rao2011}, determined using the PLC relation of RV Tauri variables in the LMC as established by \citet{Alcock_1998}. It has also been identified as an RV Tauri pulsator with a fundamental pulsation period of 40.52 days and a spectral type Fp, based on the Combined General Catalogue of Variable Stars survey by \citet{kholopov1998}. Additionally, \citet{Pollard1995} categorised V453\,Oph as an RVa type based on its reasonably regular LC variations (see Figure~\ref{fig:light_Curve:v453oph}). The high-resolution spectroscopic study conducted by \citet{Deero2005} recognised V453\,Oph as the first \sprocess enriched but carbon-deficient RV Tauri star in the Galaxy. However, the SED of this star does not exhibit any IR excess, leading to its position in the Non-IR box (see Figure~\ref{fig:SED_v453oph}). Drawing insights from the abundance analysis and metal-poor nature (\feh\ = $-2.23\pm0.12$ dex) of V453\,Oph \citep{Deero2005}, we conclude that the mild \sprocess enrichment ($\sfe\ = 0.50\pm0.21$ dex) of V453\,Oph reflects the intrinsic local \sprocess enrichment of the Galaxy at its birthplace.

\section{Dynamic Spectra of HD 158616}
\label{sec:hd158616_dynamic}

\begin{figure}
      \includegraphics[width=\columnwidth]{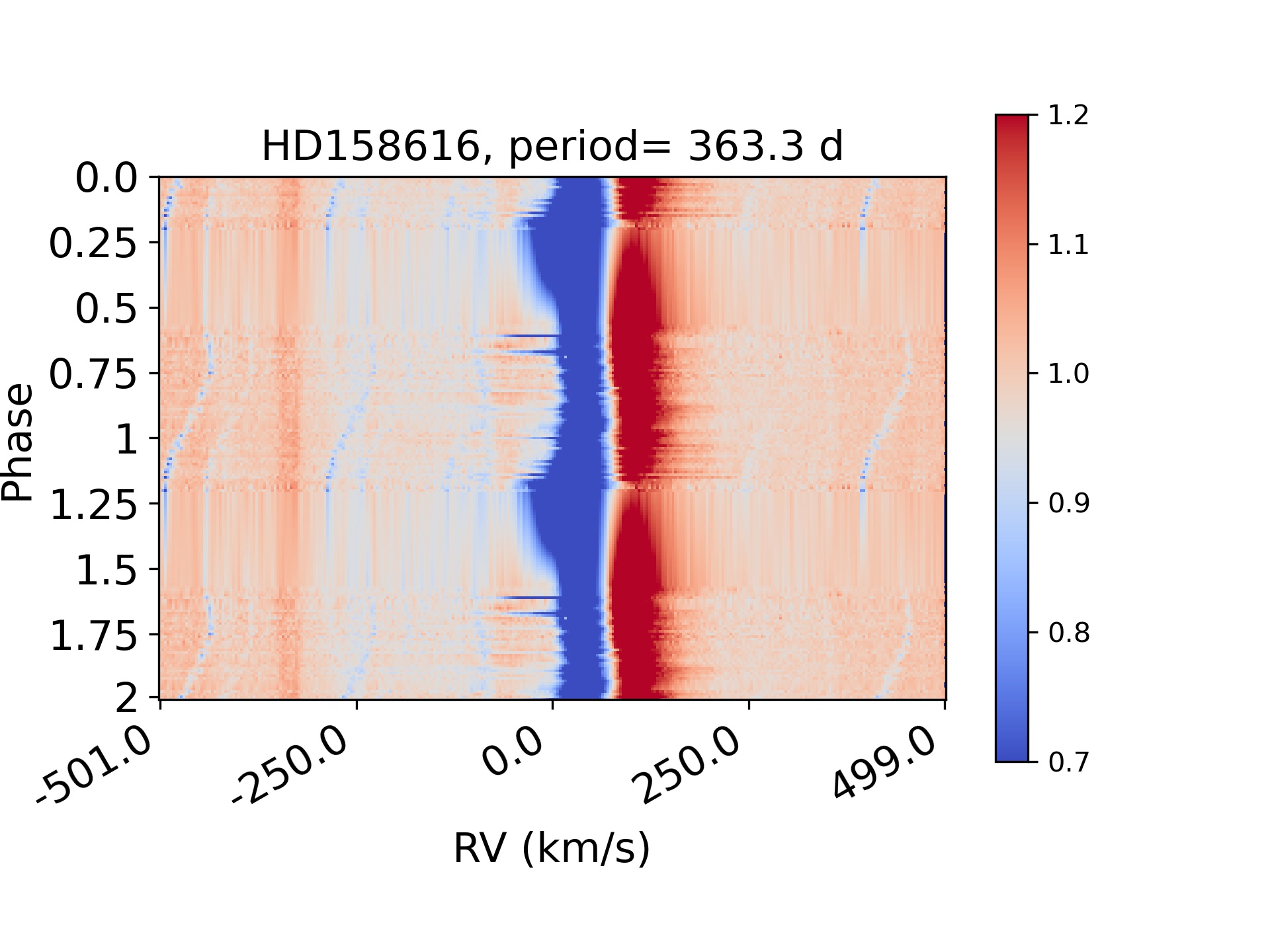}
    \caption{The dynamic spectra for the $H\alpha$ line of HD\,158616. The spectra are shown as a function of the orbital phase \citep[see][for details]{Bollen2021}.}
    \label{fig:hd158616_dynamic}
\end{figure}

\end{document}